\documentclass[12pt]{article}

\usepackage{amsfonts, amsmath, amssymb}
\usepackage{bbm}
\usepackage[dvips]{graphicx}
\usepackage{cite}

\newcommand{\bmat}{\left(\begin{array}}
\newcommand{\emat}{\end{array}\right)}

\newcommand{\pr}{\mathbbm{R}}

\def\a{\alpha}

\def\b{\beta}

\def\d{\delta}

\def\-{\hphantom{-}}

\def\s2{\frac{1}{\sqrt2}}

\def\oh{\frac{1}{2}}
\def\beq{\begin{equation}}
\def\eeq{\end{equation}}
\def\beqa{\begin{eqnarray}}
\def\eeqa{\end{eqnarray}}

\def\Tr{{\rm Tr \,}}

\def\ca{{\mathcal A}}

\def\cc{{\mathcal C}}

\def\cam{{\mathcal M}}
\def\cn{{\mathcal N}}

\def\cf{{\mathcal F}}

\def\deq#1{\mbox{$D$=#1}}
\def\neq#1{\mbox{$\cn$=#1}}
\def\Dsl{\,\raise.15ex\hbox{/}\mkern-13.5mu D} 
\def\r#1{\mbox{{\bf #1}}}

\newcommand{\mathsm}[1]{\mbox{\small$#1$}}
\def\r#1{{\bf #1}}
\def\br#1{{\bf \overline{#1}}}

\def\RR{{\pmb{R}}}
\def\rr{{\pmb{r}}}
\def\bz{{\bar z}}
\def\bw{{\bar w}}
\def\bu{{\bar u}}

\def\aa{{\bar \alpha}_a}
\def\ab{{\bar \alpha}_b}
\def\ac{{\bar \alpha}_c}
\def\db{{\bar \delta}}
\def\ba{{\bar \beta}_a}
\def\bb{{\bar \beta}_b}
\def\bc{{\bar \beta}_c}
\def\rb{{\bar \rho}}
\def\Ca{{\bar C}_a}
\def\Cb{{\bar C}_b}
\def\Cc{{\bar  C}_c}
\def\Da{{\bar D}_a}
\def\Db{{\bar D}_b}

\def\e{\epsilon}

\begin{document}
\pagestyle{plain}

\makeatletter
\@addtoreset{equation}{section}
\makeatother
\renewcommand{\theequation}{\thesection.\arabic{equation}}
\pagestyle{empty}
\rightline{ IFT-UAM/CSIC-09-31}
\vspace{0.5cm}
\begin{center}
\LARGE{ Matter wave functions and Yukawa couplings  in \\
  F-theory  Grand  Unification 
\\[10mm]}
\large{A. Font$^1$ and L.E. Ib\'a\~nez$^2$ \\[6mm]}
\small{
${}^1$  Departamento de F\'{\i}sica, Centro de F\'{\i}sica Te\'orica y Computacional \\[-0.3em]
 Facultad de Ciencias, Universidad Central de Venezuela\\[-0.3em]
 A.P. 20513, Caracas 1020-A, Venezuela\\[2mm] 
${}^2$ Departamento de F\'{\i}sica Te\'orica 
and Instituto de F\'{\i}sica Te\'orica UAM-CSIC,\\[-0.3em]
Universidad Aut\'onoma de Madrid,
Cantoblanco, 28049 Madrid, Spain 
\\[8mm]} 
\small{\bf Abstract} \\[5mm]
\end{center}
\begin{center}
\begin{minipage}[h]{15.0cm} 
We study the local structure of zero mode wave functions of chiral matter fields in F-theory unification. 
We solve the differential equations for the zero modes derived from  local Higgsing in the 8-dimensional parent action of F-theory 7-branes. 
The solutions are found as expansions both in powers and derivatives of the magnetic fluxes.
Yukawa couplings are given by an overlap integral of the
three wave functions involved in the interaction and can be calculated analytically.
We provide explicit expressions for these Yukawas to second 
order both in the flux and derivative expansions and discuss the effect of higher order terms.
We  explicitly describe  the dependence  of the couplings on the
$U(1)$ charges of the relevant fields, appropriately taking into account
their normalization.  A hierarchical Yukawa structure is naturally  obtained.  The application of our results to the
understanding of the observed hierarchies of quarks and leptons is discussed.

\end{minipage}
\end{center}
\newpage
\setcounter{page}{1}
\pagestyle{plain}
\renewcommand{\thefootnote}{\arabic{footnote}}
\setcounter{footnote}{0}

\section{Introduction}

The hierarchical structure of fermion masses and mixings is one of the most
remarkable properties of the Standard Model (SM). An outstanding
challenge of string theory compactifications is to obtain models
with the massless spectrum of the SM  and reproducing naturally such
hierarchical structure. In type IIB orientifold, as well as heterotic, 
compactifications the Yukawa couplings which govern fermion masses 
and mixings are in principle calculable, in the large compact volume limit,
in terms of overlap integrals \cite{GSW2,SW}
\beq
 Y_{ij} \ =\ \int_{X_3}\ \!\!  \psi_i \, \psi_j \, \phi_H   
\eeq
Here $\psi_{i}$  and $\phi_H$
are internal wave functions associated to the fermions and Higgs fields respectively,
taking values in the compact complex threefold $X_3$.
These wave functions are zero modes of higher dimensional fields in
the compact internal space. The technical problem here is that in general we do not
know how to compute the relevant wave functions for arbitrary curved 
spaces $X_3$. Such a computation has been completely worked out 
for the relatively simple case of type IIB toroidal orientifolds \cite{cim1} 
with constant $U(1)$ fluxes (see also \cite{cmq,vechia,antoniadis}).
In this case with a flat geometry the equations of motion can be fully solved
to obtain the wave functions which turn out to have a
neat expression in terms of Jacobi $\vartheta$-functions.
It was found that in the simplest models only one generation of quarks
and leptons acquires a non-trivial Yukawa coupling, which is 
a good starting point to reproduce the observed hierarchies \cite{cim1}.
Having explicit solutions for the wave functions  is also  useful to 
study other physical properties of the compactifications such as 
the effect of closed string fluxes and warping \cite{marchesano}.

\begin{figure}
\begin{center}
\includegraphics[scale=0.5]{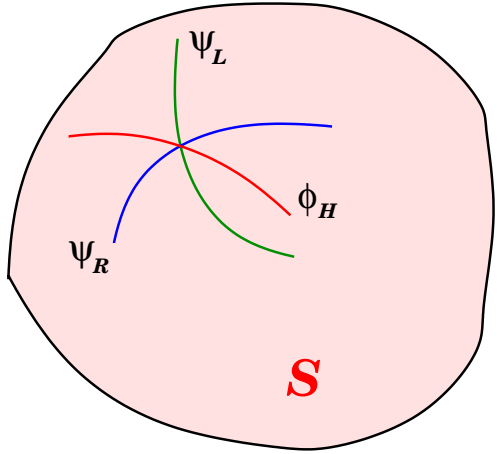}
\end{center}
\caption{\small Intersecting matter curves}
\label{cuscus}
\end{figure}

Clearly, it would be interesting to obtain wave functions and Yukawa couplings in more 
complicated curved geometries and for non-constant fluxes.  An  obvious obstruction is that
determining the wave functions seems to require a knowledge of the global
geometry of the compact $X_3$ manifold. In fact, the problem may be more tractable 
within the context of a bottom-up approach as advocated in \cite{aiqu}
(see also \cite{bjl,cgqu,vw,cmq1}). The idea is that in order to extract the 
relevant physics of a SM compactification it is enough to have a local
description of the geometry of the branes in which the SM fields reside.
This is the case for example of models derived from 
D3-branes at  singularities \cite{aiqu,bjl,vw} in which the SM physics only
depends on the local geometry around  the singularity. This 
type of structure is also characteristic of local configurations of  D7-branes wrapping intersecting 4-cycles  
inside $X_3$. F-theory \cite{ftheory} is the natural non-perturbative extension of these local 
7-brane configurations. In the last year local F-theory GUT constructions 
have been proposed \cite{dw,bhv1,bhv2} as a particularly attractive class of
bottom-up configurations with a number of phenomenological virtues (see 
\cite{w,aci,hv3,fi,hv4,mssn,hv5,ttw,blumen,rs,bhsv,bckmq,hksv,li,jlnx}, as well as
 \cite{httwy,tatar2,bourjaily,dw1,htv,dw2,ac,hmssnv,mssn2,bgjw,coll,mssn3} for other recent developments).
In this scheme the Yukawa couplings arise again as overlap 
integrals now of the form
\beq
 Y_{ij} \ =\ \int_{S}\ \psi_i \, \psi_j \, \phi_H
\label{yukayuka}
\eeq
in which $S$ is the compact complex twofold wrapped by the GUT F-theory 7-brane. The quark and lepton 
multiplets  of the SM  reside at matter Riemann curves $\Sigma_I$ 
inside $S$, which correspond geometrically to the intersection
of $S$ with the world-volume of other $U(1)$ 7-branes. Yukawa couplings 
come from the triple overlap of these matter curves involving 
quarks, leptons and Higgs fields (see figure \ref{cuscus}).
In order to compute the Yukawa coupling (\ref{yukayuka}) we need
again the internal wave functions. However, in this case given the
local geometry of the coupling it would be enough to have a
knowledge of the wave functions close to the intersection point.
It was pointed out in \cite{bhv1} that one can determine the profile of these
wave functions close to the intersection point in terms of 
a certain quasi-topological theory in \deq8. The equations of motion of
that theory have solutions  corresponding to  hypermultiplet zero modes
localized along the matter curves with a Gaussian profile.
One finds  \cite{hv3} that to leading order, for a compactification
having three generations, only one of them gets a non-trivial
Yukawa, in analogy with the results in  \cite{cim1}. However  the distortion of the wave functions, 
due to the presence of $U(1)$ gauge fluxes,  could be the natural source of the
observed hierarchy of masses and mixings of quarks and leptons \cite{hv3}.

In this article we make a systematic study of the solutions of the differential equations of motion 
of the quasi-topological \deq8 theory of \cite{bhv1}. The zero mode solutions give the local wave functions 
corresponding to the massless particles residing at the matter curves of F-theory unification models. We 
make an expansion both in powers and derivatives of the  $U(1)$ fluxes and explicitly solve the differential equations. 
Equipped with these wave functions we compute the Yukawa couplings from the overlap integral of the three wave functions 
involved in the couplings. These integrals may be calculated analytically  and we provide explicit expressions for these Yukawa couplings  
up to fourth order in the flux and derivative expansions. As suggested in \cite{hv3}, a hierarchy of masses for fermions naturally appears. 
We also study the application of our results to the understanding of the observed hierarchy of masses and mixings in the SM. We find good  
qualitative agreement with experiment for reasonable ranges of the flux parameters.

The organization of the rest of this article is as follows. In the next section we provide a
brief review of the aspects of F-theory models that concern our discussion. In chapter 3
we study the wave functions of the zero modes which are solutions of the 
quasi-topological \deq8 field theory equations. We consider both constant and varying fluxes
in a general setting  of three intersecting matter curves. The details 
of the solutions are given in appendix A. In chapter 4 we address the explicit
computation of the Yukawa couplings by evaluating the overlap integral 
of the three relevant wave functions. Based on the leading terms in the Yukawa couplings 
provided in appendix B, we describe the general structure of the flux-induced 
corrections and their contribution to the Yukawa matrices. 
In chapter 5 we apply the previous results to the analysis of the fermion mass
spectra of  $SU(5)$ GUT's with non-vanishing hypercharge flux breaking the 
theory down to the SM. We show that reasonable agreement with observed 
mass hierarchies and mixing may be obtained for appropriate flux
parameters. Chapter 6 is devoted to some final comments and discussion.

The computation of Yukawa couplings in analogous settings has been 
more recently analyzed in \cite{cchv,cp}, see 
the note added at the end of this paper.

\section{Review of F-theory unification}

The purpose of this section is to give a short overview of the F-theory formalism developed in
\cite{bhv1} (see also \cite{dw,w}).

In the F-theory setup, the \deq4 supersymmetric gauge theory descends from 7-branes wrapping a 
compact surface $S$ of complex codimension one in the threefold base of an elliptically-fibered Calabi-Yau fourfold. 
The gauge group $G_S$ on the 7-branes depends on the singularity type of the elliptic fiber. 
In turn $G_S$ can be broken by a vev in a subgroup $H_S \subset G_S$. We consider $H_S=U(1)$,  
typical examples being the hypercharge in $SU(5)$ or $U(1)_{B-L}$ in an $SO(10)$ GUT.
The $U(1)$ background breaks the gauge group and gives rise to matter charged under the commutant
of $H_S$ in $G_S$. We assume that $S$ is a del Pezzo surface so that gravity decouples from the gauge theory \cite{bhv1}.

The singularity type of the elliptic fiber can be enhanced to group $G_\Sigma$
along a curve $\Sigma \subset S$ of complex codimension
two on the threefold base. This curve appears at the intersection of $S$ and another surface $S^\prime$.
On the 7-branes wrapping $S^\prime$ there is a gauge theory with group $G_{S^\prime}$ which decouples
when $S^\prime$ is non-compact. Based on the
knowledge of intersecting D-branes, one expects additional degrees of freedom due
to open strings stretching between the 7-branes wound on $S$ and $S^\prime$. The extra fields 
localized on the matter curve $\Sigma$ must be charged under $G_S \times G_{S^\prime}$. 
Indeed this is the picture that arises in F-theory \cite{kv, bhv1}.

We will now review the basic facts about the charged fields originating in the surface $S$
and in the matter curve $\Sigma$. Our discussion is brief and follows mostly \cite{bhv1}.

\subsection{Bulk fields}
\label{ss:bulk}

The effective physics of the 7-branes wrapping $S$
is described by \deq8 twisted super Yang-Mills on $\pr^{3,1} \times S$ \cite{bhv1}.
The supersymmetric multiplets include the gauge field, plus a complex scalar $\varphi$ and fermions
$(\eta, \psi, \chi)$ in the adjoint. After twisting the scalar and fermions become forms on $S$. 
Using local coordinates $(z_1, z_2)$ for $S$ the results are summarized by
\beqa
A  =  A_\mu dx^\mu + A_m dz^m + A_{\bar m} d\bz^m \quad & ; & \quad \varphi=\varphi_{12} \, dz^1 \wedge dz^2
\nonumber \\
\psi_\a  =  \psi_{\a\bar 1} \, d\bar z^1 + \psi_{\a\bar 2} \, d\bar z^2 
\quad & ; & \quad \chi_\a=\chi_{\a 12} \, dz^1 \wedge dz^2
\label{d8fdefs}
\eeqa
Notice that $\psi$ is a (0,1) form whereas $\varphi$ and $\chi$ are (2,0) forms. The remaining
fermion $\eta_\a$ is a (0,0) form. 
The subscript $\a$, which corresponds to left handed fermions in $\pr^{3,1}$, will be dropped hereafter. 
The \deq4, \neq1  theory has gauge multiplet $(A_\mu, \eta)$, together with chiral multiplets
$(A_{\bar m}, \psi_{\bar m})$ and $(\varphi_{12}, \chi_{12})$, plus their complex conjugates.

The \deq8 effective action found in \cite{bhv1} can be integrated over the compact surface $S$ to obtain
the dynamics of the \deq4 multiplets. In computing couplings of the charged fields the most interesting term 
will be the  superpotential
\beq
W=  M_*^4\int_{S} \Tr({\pmb F_S^{(0,2)}}\wedge {\pmb \Phi}) = 
M_*^4\int_{S} \Tr({\bar \partial}{\pmb A}\wedge {\pmb \Phi})
\, + \, M_*^4 \int_S \Tr ({\pmb A} \wedge {\pmb A} \wedge {\pmb \Phi} )
\label{wdef}
\eeq  
where $M_*$ is the mass scale characteristic of the supergravity limit of F-theory. 
Here $\pmb A$ and $\pmb \Phi$ are chiral superfields with components
\beqa
{\pmb A}_{\bar m} & = & A_{\bar m}   + \sqrt2 \theta \psi_{\bar m} + \cdots \nonumber \\[2mm]
{\pmb \Phi}_{12} & = & \varphi_{12} + \sqrt2 \theta \chi_{12} + \cdots
\label{superfi}
\eeqa
where $\cdots$ involves auxiliary fields. Only the (0,2) component of the superstrength appears 
in (\ref{superfi}).

The equations of motion derived from the \deq8 effective action are the starting point to discuss the
zero modes. The part of the action bilinear in fermions, without kinetic terms,
is found to be \cite{bhv1}
\beq
I_F = \int_{\pr^{3,1}\times S} \hspace*{-.8cm} d^4x \ \Tr\big(\chi \wedge \partial_A \psi + 
2i\sqrt2 \omega \wedge\partial_A \eta \wedge \psi + \oh \psi \wedge [\varphi, \psi] + \sqrt2 \eta [\bar\varphi,\chi]
+ {\rm h.c.}\big)
\label{eq451}
\eeq 
where $\omega$ is the fundamental form of $S$. 
Taking variations with respect to $\eta $, $\psi$ and $\chi$ respectively 
gives the equations of motion
\beqa
&& \omega \wedge \partial_A \psi + \frac{i}2 [ \bar \varphi, \chi]=0  \label{zma}\\[2mm]
&& \bar \partial_A \chi - 2i\sqrt2 \omega \wedge \partial \eta - [\varphi, \psi] = 0 \label{zmb}\\[2mm]
&& \bar \partial_A \psi - \sqrt2 [\bar \varphi, \eta] = 0 \label{zmc}
\eeqa 
For the bosonic fields it is found that
the field strength $F_S$ must have vanishing (2,0) and (0,2) components and verify the BPS condition
\beq
\omega \wedge F_S + \frac{i}2 [\varphi, \bar \varphi] = 0
\label{bpscond}
\eeq
Finally, the complex scalar must satisfy the holomorphicity condition 
$\bar\partial_A \varphi=0$. 

To determine the charged massless multiplets in \deq4 it is necessary to specify the background for 
the adjoint scalar $\varphi$ and the gauge field. When $\langle \varphi \rangle=0$, the equations of motion 
imply that the number of zero modes of $\psi$ and $\chi$ are counted by topological invariants that depend both 
on $S$ and the gauge bundle of the background \cite{bhv1}.

\subsection{Fields at intersections}
\label{ss:sigma}

We now want to discuss the degrees of freedom localized on a matter curve $\Sigma$
occurring at the intersection of surfaces $S$ and $S^\prime$. As explained in \cite{bhv1},
to preserve \neq1 supersymmetry in \deq4, the theory on $\pr^{3,1} \times \Sigma$ must be
\deq6 twisted super Yang-Mills. The \deq6 twisted supermultiplet, which includes two complex
scalars and a Weyl spinor, decomposes into \deq4 chiral multiplets $(\sigma, \lambda)$ and
$(\sigma^c, \lambda^c)$, plus CPT conjugates. The number of zero modes is given by topological
invariants depending on $\Sigma$ and the gauge bundle on a background in $G_\Sigma$.   

There is a very nice intuitive way of understanding the matter localized at $\Sigma$. The idea,
originally given in \cite{kv} and expanded in \cite{bhv1}, is to start from the \deq8 theory on $S$ with 
gauge group $G_\Sigma$ and then turn on a background for the adjoint scalar given by
\beq
\langle \varphi \rangle = m^2 z_1 Q_1 
\label{vev1}
\eeq
where $z_1$ is a complex coordinate on $S$, and $Q_1$ is a $U(1)_1$ generator in the Cartan subalgebra
of $G_\Sigma$. To streamline notation, $\varphi = \varphi_{12}$.
We have explicitly introduced a mass parameter $m$ so that $\varphi$ has the standard dimensions.
The basic idea is that in presence of $\langle \varphi \rangle$ the \deq8 fields have zero
modes localized at $z_1=0$ that are naturally associated to the fields at the intersection.

When $z_1=0$ the gauge group is unbroken, but when $z_1 \not=0$ the group is broken to $G_S \times U(1)_1$,
with $G_S$ being the group whose generators commute with $Q_1$. The locus $z_1=0$ defines the curve $\Sigma$.
Thus, on $\Sigma$ the singularity enhances from $G_S$ to $G_\Sigma \supset G_S \times U(1)_1$. The breaking
of the gauge group is explained by the deformation of the singularity type from $G_\Sigma$ to $G_S$ due to the
background in the Cartan subalgebra \cite{kv}.

For ordinary D7-branes the adjoint scalar corresponds to degrees of freedom in the transverse direction and a
non-zero vev means that some branes are separated and the gauge group is broken. For instance, if 
there are $(K+1)$ D7-branes to begin and one is moved away, the original $SU(K+1)$ is broken to $SU(K) \times U(1)$.
Furthermore, the open strings stretching between the two stacks of D7-branes give rise to massless bifundamentals
$({\pmb K}, -1) + (\overline{\pmb K}, 1)$ localized at the intersection. For F-theory seven-branes wrapping
a surface $S$ one then expects  $\langle \varphi\rangle$ to break the original gauge group to some $G_S$. Moreover,
there will be massless `bifundamentals' descending from the adjoint of $G_\Sigma$ which decomposes as a direct sum
of irreducible representations $(\RR, q_1)$ under  $G_S \times U(1)_1$.

Several examples of singularity resolution were worked out in \cite{kv} and more recently in
\cite{bhv1, bhv2, bourjaily}. For illustration let us consider the case $G_\Sigma=E_6$ and
$G_S=SO(10)$ that will be of interest later on. Under $SO(10) \times U(1)$ the $E_6$ adjoint decomposes as
\beq
\r{78}   =  (\r{45},0) + (\r1,0) + (\r{16},1) + (\br{16},-1)    
\label{e6br}
\eeq
Therefore, there will be chiral multiplets transforming as $\r{16}$ and $\br{16}$ of $SO(10)$. To see how
$SO(10)$ is enhanced to $E_6$ on $\Sigma$ it is convenient to represent the Cartan generators as vectors 
$|Q_i\rangle$ so that $|\varphi \rangle \propto z_1 | Q_1\rangle$ corresponds to the adjoint vev \cite{kv}.
The simple roots are elements $\langle v_j|$ of the dual space. Those roots with $\langle v_j | \varphi\rangle=0$
remain as $SO(10)$ roots while those with $\langle v_j | \varphi\rangle \propto z_1$ become the weights of
the $\r{16}$ and $\br{16}$.      

The resolution of the singularity by the adjoint vev can be figured out as explained in \cite{kv}.
The generic $E_6$ singularity can be cast as \cite{km, bhv1}
\beq
y^2 = x^3 + \frac14 z^4 + \e_2 x z^2 + \e_5 xz + \e_6 z^2 + \e_8 x + \e_9 z + \e_{12}
\label{e6sing}
\eeq
where the $\e_i$ are functions that depend on the adjoint vev. More precisely, in \cite{km} they are given in terms
of an arbitrary vector $(t_1, \cdots, t_6)$ in the $E_6$ Cartan subalgebra. In our notation, in the
$E_6 \supset SO(10) \times U(1)_1$ example, $t_1=z_1$ while other $t_i$'s vanish. By choosing $t_1=-3t$, and computing 
the $\e_i$ according to the formulas of \cite{km}, we obtain the deformation 
\beq
y^2 =x^3 + \frac14 z^4 - 3 t^2 x z^2 - 12 t^5  xz - 6 t^6 z^2 - 12 t^8 x - 16 t^9 z - 12 t^{12}
\label{so10sing}
\eeq 
This is the same result found in \cite{kv}, for a different though equivalent choice of vev vector. 
It can be shown that for $t\not=0$ there is an $SO(10)$ singularity. 

So far we have just reviewed how the gauge group on the curve $\Sigma$ is enhanced. We now want 
to discuss how the matter localized on $\Sigma$ arises from zero modes of the \deq8 bulk fields. 
It is enough to look at fermions because the scalars follow by supersymmetry. We then want
to solve the \deq8 equations of motion for the twisted fermions when $\varphi$ has the vev
linear in $z_1$, and there is no gauge background. The solutions that are localized at $z_1=0$ can be interpreted 
as the fermions $\lambda$ and $\lambda^c$ that come from the twisted super Yang-Mills on    
$\pr^{3,1} \times \Sigma$. 

We start from the \deq8 fermionic equations of motion (\ref{zma}-\ref{zmc}).
To show that there are localized solutions it suffices to work locally 
and assume that the fundamental form of $S$ has the canonical Euclidean form 
\beq
\omega = \frac{i}2 (dz^1 \wedge d \bz^1 + dz^2 \wedge d \bz^2)
\label{oms}
\eeq
Notice that the coordinate along $\Sigma$ is $z_2$ whereas the transverse coordinate is $z_1$. To
look for localized solutions one can neglect derivatives in $z_2$. The equations of motion reduce then to
\beqa
\sqrt2 \, \partial_1 \eta - m^2 z_1 q_1 \psi_{\bar 2} = 0 \quad & ; & \quad
\bar \partial_{\bar 1} \psi_{\bar 2} - \sqrt2 \, m^2 \bz_1 q_1\eta  = 0  \label{etapsi2} \\[2mm]
\partial_1 \psi_{\bar 1} - m^2 \bz_1 q_1 \chi = 0 \quad & ; & \quad
\bar \partial_{\bar 1} \chi - m^2 z_1 q_1 \psi_{\bar 1} \label{chipsi1} = 0
\eeqa   
where $\chi=\chi_{12}$. Here $q_1$ is the $U(1)_1$ charge of the fermions that belong to a
representation $(\RR, q_1)$ of  $G_S \times U(1)_1$. {}From the above equations we see that
there are no localized solutions for $\eta$ and $\psi_{\bar 2}$, and indeed it is consistent to
set $\eta=0$ and $\psi_{\bar2 }=0$. On the other hand, the coupled system for $\chi$ and
$\psi_{\bar 1}$ has solution
\beq
\chi=f(z_2) \, e^{-e m^2 |z_1|^2} \quad ; \quad 
\psi_{\bar 1} = - f(z_2) \, e^{-e m^2 |z_1|^2} 
\label{locsol}
\eeq   
where $f(z_2)$ is an arbitrary holomorphic function of the coordinate along $\Sigma$.
We have set $q_1=e$ to take into account normalization of the charges.
Clearly the zero modes are peaked around $z_1=0$, with width in $|z_1|^2$ equal to $1/e m^2$.
The constant $em^2$ is expected to be of the order of the F-theory  mass scale $M_*^2$. 

The solutions localized at $z_1=0$  naturally correspond to the fermions
$\lambda$ and $\lambda^c$ that appear in the \deq6 twisted theory. As argued in \cite{bhv1}, the 
transformations along $\Sigma$ of $(\psi_{\bar 1}, \chi)$ and $(\lambda, \lambda^c)$ do agree.

\section{Zero modes at intersecting matter curves}
\label{s:local}

As we have reviewed, there are charged fields localized on a matter curve $\Sigma$ 
where the singularity type is enhanced. In this section we want to study the situation
in which there are three matter curves $\Sigma_I$, $I=a,b,c$,  occurring at the intersection of surfaces $S$ and $S^\prime_I$.
In turn the three matter curves intersect at a point.
On each curve there is a gauge group $G_{\Sigma_I}$ that enhances to $G_p$ at the common
point of intersection \cite{bhv1}.

To obtain the wave functions of the fermionic zero modes living at intersections we follow again the approach of \cite{bhv1}.
The strategy is to consider the fermionic equations of motion of the \deq8 theory on $S$ with a non-trivial 
background for the adjoint scalar $\varphi$ that determines the curves. One then looks for solutions that are localized 
on a particular matter curve.

In the previous section we have seen that the equations that can give rise to localized fermionic zero modes 
are given by 
\beq
\omega \wedge \partial_A \psi + \frac{i}2 [\langle \bar \varphi \rangle, \chi] = 0 \qquad ; \qquad
\bar \partial_A \chi -[\langle \varphi \rangle, \psi] = 0
\label{zm1}
\eeq
We have set $\eta=0$ because only fermions that appear in \deq4 chiral multiplets are expected to
have localized modes on $\Sigma$. Notice then that equation (\ref{zmc}) implies the additional
constraint $\bar \partial_A \psi=0$. 
The new ingredient now is a more general background for the adjoint scalar $\varphi$. Concretely, 
\beq
\langle \varphi \rangle = m_1^2 \, z_1 \, Q_1 + m_2^2 \, z_2 \, Q_2
\label{phibg}
\eeq
where the $z_i$ are local coordinates, and $\varphi=\varphi_{12}$.
Each $Q_i$ is the generator of a $U(1)_i$ inside $G_p$.
The $m_i$ are some mass parameters expected to be related to the F-theory supergravity  scale $M_*$.
In what follows we  will take  $m_1=m_2=m$. 

As discussed in section \ref{ss:sigma}, when $\langle \varphi \rangle \propto z_1$, the adjoint vev
resolves the singularity on the curve $\Sigma_a$ characterized by $z_1=0$ \cite{kv}. Now the  
more general adjoint background resolves the $G_p$ singularity where three curves intersect.     
When $(z_1,z_2) \not= (0,0)$, the group is broken to $G_S$ but at the intersection $(z_1,z_2) = (0,0)$ the group
is enhanced to $G_p$. Furthermore, at the curves $\Sigma_I$ the group is enhanced to $G_{\Sigma_I} \supset G_S \times U(1)$.
For example, when $(G_p, G_S)=(E_7, SO(10))$, the group is enhanced to $E_6$ at $\Sigma_a$ and $\Sigma_b$ defined 
respectively by the loci $z_1=0$ and $z_2=0$, whereas it is enhanced to $SO(12)$ at $\Sigma_c$ defined by $z_1+z_2=0$.
In the case  $(G_p, G_S)=(E_8, E_6)$, the group is enhanced to $E_7$ at each curve. 

At each curve $\Sigma_I$ there are matter fermions that correspond to open strings 
stretching between 7-branes wrapping $S$ and $S^\prime_I$. The $U(1)_i$ charges of these fermions, denoted $(q_1,q_2)$, 
depend on the curve as shown in table \ref{sigmaq}. In this table we also indicate how the fermions transform under 
the gauge group in the examples $G_S=E_6$ and $G_S=SO(10)$ in which the group $G_p$ is respectively $E_8$ and $E_7$.
For $G_S=SU(5)$ the rank two higher $G_p$ can be either $E_6$ or $SO(12)$. 
We have introduced parameters $(e_1,e_2)$ to take into account normalization of the charges.

\begin{table}[htb]
\begin{center}
\begin{tabular}{|c|c|c|c|c|c|c|}
\hline
curve & $(q_1,q_2)$ & locus & $E_6$  & $SO(10)$  & $SU(5)$ & $SU(5)$  \\
\hline\hline
 $\Sigma_a$ &  $(e_1,0)$ &  $z_1=0$ & $\r{27}$ & $\r{16}$ & $\r{10}$ & $\r{10}$ \\
\hline
 $\Sigma_b$ & $(0,e_2)$ &  $z_2=0$ & $\r{27}$ & $\r{16}$ &  $\r{10}$ & $\r{\bar 5}$ \\
\hline
 $\Sigma_c$ & $(-e_1,-e_2)$ &  $z_1+z_2=0$ & $\r{27}$ & $\r{10}$ &  $\r{5}$ & $\r{\bar 5}$ \\
\hline
\end{tabular}
\end{center} 
\caption{Curves and charges}
\label{sigmaq}
\end{table}

\subsection{Zero modes in the absence of fluxes}
\label{ss:noflux}

We will first solve the zero mode equations without turning on a background gauge field but
with scalar vev $\langle \varphi\rangle$ given in (\ref{phibg}). 
The fundamental form $\omega$ is assumed to have the standard local form (\ref{oms}).
Recall that $\psi=\psi_{\bar \imath} d\bz^i$ and $\chi=\chi_{12}dz^1 \wedge dz^2$ are forms on $S$. 
Substituting in the master equations (\ref{zm1}) yields
\beqa
\partial_2 \psi_{\bar 2} +  \partial_1 \psi_{\bar 1} - m^2(\bz_1 q_1 + \bz_2 q_2) \, \chi & = & 0 \nonumber \\[1mm]
\bar \partial_{\bar 1} \chi - m^2(z_1 q_1 + z_2 q_2) \, \psi_{\bar 1} & = & 0 \label{noflux} \\[1mm]
\bar \partial_{\bar 2} \chi - m^2(z_1 q_1 + z_2 q_2) \, \psi_{\bar 2} & = & 0 \nonumber
\eeqa
where now $\chi=\chi_{12}$.
The constants $(q_1,q_2)$ are the $U(1)_i$'s charges of the fermions that belong to a
representation $(\RR, q_1, q_2)$ of  $G_S \times U(1)_1 \times U(1)_2$.
In the following we will analyze the different possibilities for the fermions 
with charges and corresponding curves shown in table \ref{sigmaq}.
Notice that the condition $\bar\partial_{\bar 2} \psi_{\bar 1}= \bar\partial_{\bar 1} \psi_{\bar 2}$, implied by
the additional constraint $\bar \partial \psi=0$, is automatic by virtue of the last two equations above.

\bigskip
\noindent
\underline{$\Sigma_a$, $(q_1,q_2)=(e_1,0)$}

\medskip
\noindent
After substituting the charges in (\ref{noflux}) we obtain the solutions
\beq
\psi_{\bar 2} = 0 \quad ; \quad 
\chi=f(z_2) \, e^{-\lambda_1 |z_1|^2} \quad ; \quad 
\psi_{\bar 1} = -\frac{\lambda_1}{e_1m^2} \, \chi 
\label{s1nf}
\eeq
where $f(z_2)$ is a holomorphic function of $z_2$. The equations (\ref{noflux}) require the constant $\lambda_1$
to satisfy
\beq
\lambda_1^2 = e_1^2 m^4
\label{l1sol}
\eeq
We see that there are solutions localized at $z_1=0$ provided that 
we take the positive root $\lambda_1=e_1 m^2$.  
We then have two zero modes $\psi_{\bar 1}$ and $\chi$ which correspond to
massless fermions of $D=6$ massless hypermultiplets living on $\Sigma_a$.
In the presence of magnetic fluxes through $\Sigma_a$, chiral four dimensional
fermions will appear coming from $\psi_{\bar 1}$  or/and $\chi $ as dictated by
index theorems.

The characteristic width of the Gaussian wave functions is $v=1/(e_1m^2)$. We will assume that $v=1/M_*^2$,
where $M_*$ is the F-theory mass scale. For sufficiently large compactification
radius $R$ this width becomes negligibly small.

\bigskip
\noindent
\underline{$\Sigma_b$, $(q_1,q_2)=(0,e_2)$}

\medskip
\noindent
In this case the solutions of (\ref{noflux}) turn out to be 
\beq
\psi_{\bar 1} = 0 \quad ; \quad \chi=g(z_1) \, e^{-\lambda_2 |z_2|^2} \quad ; \quad 
\psi_{\bar 2} = -\frac{\lambda_2}{e_2m^2} \, \chi 
\label{s2nf}
\eeq
with $g(z_1)$ a holomorphic function of the longitudinal coordinate $z_1$. The constant $\lambda_2$
now satisfies
\beq
\lambda_2^2 = e_2^2 m^4
\label{l2sol}
\eeq
As in the previous situation, having solutions localized at $z_2=0$ requires $\lambda_2=e_2 m^2$.

\bigskip
\noindent
\underline{$\Sigma_c$, $(q_1,q_2)=(-e_1,-e_2)$}

\medskip
\noindent
To treat this curve it is convenient to introduce new variables and fields, and to simplify by setting $e_1=e_2=e$.
Consider then the definitions
\beqa
w=z_1 + z_2 \quad & ; & \quad \psi_{\bar w} = \oh (\psi_{\bar 1} + \psi_{\bar 2}) \nonumber \\[1mm]
u=z_1 - z_2 \quad & ; & \quad \psi_{\bar u} = \oh (\psi_{\bar 1} - \psi_{\bar 2}) \label{newdefs}
\eeqa
The zero mode equations then become
\beqa
2(\partial_w \psi_{\bar w} + \partial_u \psi_{\bar u}) + em^2\bar w \, \chi & = & 0 \nonumber \\[1mm]
\bar \partial_{\bar w} \chi +   em^2 w\,  \psi_{\bar w} & = & 0 \label{noflux3} \\[1mm]
\bar \partial_{\bar u} \chi  +  em^2 w \, \psi_{\bar u} & = & 0 \nonumber
\eeqa
Now there are localized solutions at $w=0$, namely 
\beq
\psi_{\bar u} = 0 \quad ; \quad \chi=h(u) \, e^{-\lambda_3 |w|^2} \quad ; \quad 
\psi_{\bar w} = \frac{\lambda_3}{em^2} \, \chi 
\label{s3nf}
\eeq
with $h(u)$ a holomorphic function of the coordinate $u$ along $\Sigma_c$, and $\lambda_3=em^2/\sqrt2$.
Notice that $\psi_{\bar u}=0$ implies
\beq
\psi_{\bar 1} = \psi_{\bar 2} = \frac {1}{\sqrt{2}} \, h(u) \, e^{-\lambda_3 |w|^2}  
\label{p12s3nf}
\eeq 
These agree with results in  \cite{tatar2}.

\subsection{Zero modes with fluxes}
\label{ss:withflux}

We now want to solve the zero mode equations with a background flux, still keeping the adjoint vev
$\langle \varphi \rangle$ given in (\ref{phibg}). We already know that without flux each curve
$\Sigma_I$ supports localized modes with $U(1)_i$ charges given in table \ref{sigmaq}.
The fermions on each curve will now feel a total flux $\cf$ that includes various contributions.
There is a bulk $U(1)$ flux $F$ in $G_S$ with generator $Q$ (for example, 
hypercharge or $Q_{B-L}$). There
are also fluxes $F^{(i)}$ along the $U(1)_i$ inside $G_p$ with generators $Q_i$. The total flux can
then be written as
\beq
\cf = F\, Q + F^{(1)}\, Q_1 + F^{(2)}\, Q_2 
\label{tflux}
\eeq
The corresponding gauge potentials will be denoted $\ca$, $A$ and $A^{(i)}$, with the total potential
$\ca$ decomposed like the total flux. We will use conventions 
in which the covariant derivative $\partial_A$ of a field of charge $q$ is defined as
\beq
\partial_A = \partial - iq\, A
\label{covdef}
\eeq
All field strengths and gauge potentials are taken to be real. 

The fermions $\chi$ and $\psi$ have $U(1)_i$ charges $(q_1,q_2)$ and transform in some representation
$\RR$ of $G_S$. The bulk flux break $G_S$ to $\Gamma_S \times U(1)$ and $\RR$ decomposes into a
direct sum of irreducible representations that can be labelled by $(\rr, q, q_1, q_2)$, where $q$ is
the bulk $U(1)$ charge. The zero mode equations for the charged fermions then become
\beqa
(\partial_2 - i \ca_2) \psi_{\bar 2} + (\partial_1 -i\ca_1)\psi_{\bar 1} - 
m^2(\bz_1 q_1 + \bz_2 q_2) \, \chi & = & 0 \nonumber \\[1mm]
(\bar \partial_{\bar 1}- i\ca_{\bar 1}) \chi - m^2(z_1 q_1 + z_2 q_2) \, \psi_{\bar 1} & = & 0 \label{withflux} \\[1mm]
(\bar \partial_{\bar 2} -i\ca_{\bar 2}) \chi - m^2(z_1 q_1 + z_2 q_2) \, \psi_{\bar 2} & = & 0 \nonumber
\eeqa    
Clearly, the total gauge potential that appears depends on the charges. It is explicitly given by
\beq
\ca = q\, A + q_1 \, A^{(1)} + q_2\, A^{(2)} 
\label{tpot}
\eeq 
The task is to solve the above equations for particular fluxes. 

The 8-dimensional equations of motion further require the vanishing of the $(2,0)$ and $(0,2)$ components
of the field strengths. We will only consider diagonal components $\cf_{1\bar 1}$ and $\cf_{2\bar 2}$, 
even though $\cf_{1\bar 2}$ and $\cf_{1 \bar 2}$ are also allowed. Using local coordinates the bulk flux 
takes the form
\beq
F = F_{1\bar 1} \, dz_1 \wedge d\bz_1 +  F_{2\bar 2} \, dz_2 \wedge d\bz_2
\label{bfluxgen}
\eeq
For the $U(1)_i$ fluxes we instead take
\beq
F^{(1)}  = F^{(1)}_{2\bar 2}  \, dz_2 \wedge d\bz_2 \qquad ; \qquad 
F^{(2)}  =  F^{(2)}_{1\bar 1}  \, dz_1 \wedge d\bz_1 
\label{cfluxgen}
\eeq  
The rationale is that, say $F^{(1)}$, is the flux along the curve $\Sigma_a$ that is defined by
$z_1=0$ and has coordinate $z_2$.

We will start by analyzing constant field strengths in section \ref{sss:cflux}. In this case it is possible to
solve the zero mode equations exactly. We will then study variable fluxes that turn out to be necessary to 
generate corrections to Yukawa couplings \cite{hv3}.

\subsubsection{Zero modes with constant flux}
\label{sss:cflux}

In the case of constant field strengths the bulk flux can be written as
\beq
F = 2i M \, dz_1 \wedge d\bz_1 +  2i N \, dz_2 \wedge d\bz_2
\label{bflux}
\eeq
where $M$ and $N$ are real constants. As explained before, the $U(1)_i$ fluxes have components only 
along the curves. They are then given by
\beq
F^{(1)}  =  2i N^{(1)} \, dz_2 \wedge d\bz_2 \qquad ; \qquad F^{(2)}  =  2i M^{(2)} \, dz_1 \wedge d\bz_1 
\label{cflux}
\eeq
with $N^{(1)}$ and $M^{(2)}$ some real constants.  

For the gauge potentials we take the following gauge
\beqa
A & = &  iM \, (z_1 d\bz_1 - \bz_1 dz_1) + iN \, (z_2 d\bz_2 - \bz_2 dz_2) \nonumber \\[1mm] 
A^{(1)} & = &  iN^{(1)} \, (z_2 d\bz_2 - \bz_2 dz_2) \label{sympots} \\[1mm] 
A^{(2)} & = & iM^{(2)} \, (z_1 d\bz_1 - \bz_1 dz_1) \nonumber
\eeqa 
Notice that the total gauge potential defined in (\ref{tpot}) can be cast as
\beq
\ca =   i\cam \, (z_1 d\bz_1 - \bz_1 dz_1) + i\cn \, (z_2 d\bz_2 - \bz_2 dz_2) 
\label{tpotc}
\eeq
where the total flux coefficients are given by
\beq
\cam=(q M + q_2 M^{(2)}) \qquad ; \qquad \cn=(q N + q_1 N^{(1)})
\label{tfluxcoef}
\eeq
where $q$ and $q_i$ are the bulk and $U(1)_i$ charges respectively. 
In appendix A we  give the exact solution of the zero mode equations (\ref{withflux}) with this total 
gauge potential for the three matter curves $\Sigma_a$, $\Sigma_b$ and $\Sigma_c$.
Using these results we can then describe the localized wave functions at 
each curve.  

As explained in appendix A, it is convenient to perform a gauge transformation  \mbox{$\ca=\hat \ca + d\Omega$} such that
$\hat \ca_{\bar 1}=\hat \ca_{\bar 2}=0$, and then work with the potential $\hat \ca$.
We will refer to this choice as the holomorphic gauge.  
The wave functions in this gauge, denoted $\hat \chi$ and $\hat \psi_{\bar \imath}$, 
take a simpler form and are better suited to compute gauge invariant quantities such
as Yukawa couplings.

In the case of $\Sigma_a$ we find wave functions
\beq
\hat \chi = f(z_2)\, e^{-\lambda_1 |z_1|^2}
\quad ; \quad \hat \psi_{\bar 1} = - \frac{\lambda_1}{e_1m^2} \hat \chi
\quad ; \quad \hat \psi_{\bar 2} = 0
\label{casosigma1}
\eeq
where 
\beq
\lambda_1 = -\cam + e_1m^2\, \sqrt{1 + \frac{\cam^2}{e_1^2m^4}} =
-\cam + e_1m^2 +  \frac12 \frac{\cam^2}{e_1m^2}  + \cdots
\label{ellambda1}
\eeq
which reduces to $\lambda_1=e_1 m^2$ when $\cam =0$. For future purposes we record the expansion
of the zero modes to first order in $\cam$, namely
\beq
\hat \chi =  \hat  \chi^{(0)} \big\{ 1 + \cam |z_1|^2 + \cdots \big\} \quad ; \quad
\hat \psi_{\bar 1} =  \hat  \psi_{\bar 1}^{(0)} 
\big\{ 1 -  \cam v + \cam |z_1|^2 + \cdots \big\}
\label{chipsi10}
\eeq
where $v=1/(e_1m^2)$. Clearly, $\hat  \chi^{(0)}= - \hat  \psi_{\bar 1}^{(0)} =f(z_2)\, e^{-|z_1|^2/v}$ is
the solution for $\cam=0$.

Notice that as expected the flux has the effect 
of deforming the wave function. In the 
holomorphic gauge defined above the wave functions depend on fluxes only through $\cam$.
Since the matter fields in the curve $\Sigma_a$ have $q_2=0$ the wave function depends only on the 
flux in the bulk (e.g. from hypercharge in $SU(5)$ or $U(1)_{B-L}$ in $SO(10)$). Concretely, we must replace
$\cam$ above by
\beq
\cam_a = q_a M
\label{ema}
\eeq
where $q_a$ is the bulk charge and $M$ comes from the bulk flux. 
This is relevant later when extracting the $U(1)$ charge dependence of the Yukawa couplings.  

Analogous results are obtained for the $\Sigma_b$ matter curve
with the obvious replacements $\cam \rightarrow \cn$ and
$e_1\rightarrow e_2$. In the holomorphic  gauge the $\Sigma_b$ wave function depends
only on the bulk flux. This means that $\cn$ must be replaced by $\cn=q_b N$. 

For the $\Sigma_c$ curve the wave functions are found to be 
\beq
\hat \chi = h(u+\gamma w) \, e^{-\lambda_3 |w|^2}\, e^{\xi w \bar u}
\qquad ; \qquad \hat \psi_{\bar w} = \frac{\lambda_3}{em^2} \hat \chi \qquad ; \qquad
\hat \psi_{\bar u} = - \frac{\xi}{em^2} \hat \chi
\label{casosigma3}
\eeq
where $h(u+\gamma w)$ is an holomorphic function of its argument and
\beq
\gamma=\frac{\xi}{\lambda_3} \qquad ; \qquad \xi=\frac{\lambda_3(\cam-\Delta)}{(\lambda_3 + \Delta)}
\label{xiepstexto}
\eeq
where $\Delta=(\cam +\cn )/2$ and $\lambda_3$ is given in 
appendix A. In the absence of fluxes one has
 $\xi=\gamma=0$, and $\lambda_3=em^2/\sqrt{2}$, recovering the fluxless result.
Note that now it is the linear combination $\xi {\hat \psi_{\bar w}}
+\lambda_3\hat \psi_{\bar u}$ which vanishes.
On the curve $\Sigma_c$ the matter fields have $U(1)_i$ charges
$(q_1,q_2)=(-e_1,-e_2)$. Hence, $\cam$ and $\cn$ in this case are explicitly given by
\beq
\cam_c = q_c M - e_2 M^{(2)} \quad ; \quad \cn_c = q_c N - e_1 N^{(1)}
\label{emenc}
\eeq 
We see that the wave function depends on both bulk and brane fluxes.

\subsection{Zero modes with variable fluxes}
\label{sss:vflux}

In general it is quite complicated to obtain the exact wave functions for non-constant field strengths. 
In \cite{hv3} an adiabatic hypothesis is assumed whereby the wave functions 
basically follow from those of constant fluxes by replacing the constant coefficients $\cf_{i\bar j}$ by their variable 
counterparts. An expansion in powers of the $z_i$'s is then performed. In this article our approach
will be to consider variable fluxes expanded in powers of the local coordinates from the beginning,
and then solve the differential equations for the zero modes.

We will first expand the fields strengths up to second order in the local coordinates.
We again turn on only components $\cf_{1\bar 1}$ and $\cf_{2 \bar 2}$. Specifically, we take
\beqa
\cf_{1\bar 1} & = & 2i\cam + 4i(\a_1 z_1 + \bar\a_1 \bz_1) \, + \, 6i(\b_1 z_1^2 + \bar\b_1 \bz_1^2) 
\nonumber \\[1mm]  
\cf_{2\bar 2} & = & 2i\cn \, + \, 4i(\a_2 z_2 + \bar\a_2 \bz_2) \, + \, 6i(\b_2 z_2^2 + \bar\b_2 \bz_2^2) 
\label{vflux}
\eeqa
where the flux coefficients $\a_i$ and $\b_i$ are complex constants while $\cam$ and $\cn$ are real.
In practice the expansion parameter is $z_i/R$, where $R$ is the compactification radius (see section \ref{ss:eflux}).
We have neglected quadratic terms proportional to $(z_i\bz_i)$ because they do not give any new effects
concerning Yukawa couplings. 
The total flux coefficients can be split into bulk and curve contributions in analogy to (\ref{tfluxcoef}).

In our gauge choice the vector potential has components
\beqa
\ca_1 & = &  -i\cam \bz_1 \, - \, i(\bar\a_1\bz_1^2 + 2\a_1z_1\bz_1) \, - \, 
i(\bar\b_1 \bz_1^3 + 3\b_1\bz_1 z_1^2) \nonumber\\[1mm]
\ca_2 & = &  -i\cn \bz_2 \, - \, i(\bar\a_2\bz_2^2 + 2\a_2z_2\bz_2) \, - \, 
i(\bar\b_2 \bz_2^3 + 3\b_2\bz_2 z_2^2) 
\label{vpot}
\eeqa
whereas $\ca_{\bar 1}=\ca_1^*$, and $\ca_{\bar 2}=\ca_2^*$. In appendix A we  discuss the solutions of the
zero mode equations (\ref{withflux}) with this total gauge potential. 

We have not solved the zero mode equations exactly. Instead we found solutions in a perturbative expansion
in the flux parameters $(\cam,\cn, \a_i, \b_i)$. We first go to the holomorphic gauge with  
$\hat \ca_{\bar 1}=\hat \ca_{\bar 2}=0$, and then iterate to obtain 
$\hat\chi = \sum_{I=0} \hat \chi^{(I)}$, where $\hat  \chi^{(I)}$ is of order $I$ in the flux coefficients. 
The zeroth order wave function $\hat \chi^{(0)}$ is the fluxless solution derived in section \ref{ss:noflux}.
Once $\hat \chi$ is determined it is straightforward to deduce the $\hat \psi_{\bar \imath}$.
For example, in $\Sigma_a$, $\hat \psi_{\bar 1}=\bar\partial_{\bar 1}\hat\chi/(e_1 m^2 z_1)$, and
$\hat \psi_{\bar 2}=0$.

The iteration can be carried out to any desired order, but the number of terms will clearly be increasingly
larger. In appendix A we only display results at most up to second order in the flux parameters. Already at first
order there is an interesting feature that deserves further elaboration. To simplify the argument we set
$\b_i=0$. Then, the wave functions in the curve $\Sigma_a$ are found to be
\beqa
\hat \chi & = &  \hat  \chi^{(0)} \bigg\{ 1 + \frac43 v\a_1z_1 + \cam |z_1|^2 
+\frac23|z_1|^2 (\bar \a_1 \bz_1 + 2\a_1 z_1)  + \cdots \bigg\} \nonumber \\
\hat \psi_{\bar 1} & = &  \hat  \psi_{\bar 1}^{(0)} 
\big\{ 1 -  \cam v - \frac43 v{\bar \a_1}\bz_1 +  \cam |z_1|^2  
+\frac23|z_1|^2 (\bar \a_1 \bz_1 + 2\a_1 z_1) + \cdots \bigg\}
\label{chipsi11}
\eeqa 
where $v=1/(e_1m^2)$. For constant $\cf_{i\bar\jmath}$  we have derived the exact solutions whose expansion
to first order in $\cam$ agrees with the above results setting $\a_i=0$.  

One  point we wish to make is that in the presence of variable field strength $\cf$ the solution is not merely
obtained by adiabatically including the coordinate dependence in $\cf$. In our case this
would correspond to substituting $\cam$ by the effective value
\beq
\cam_{eff}= \cam + 2(\bar \a_1 \bz_1 + \a_1 z_1)
\label{adiabsubs}
\eeq 
Indeed, once we replace $\cam$ by $\cam_{eff}$ in the solutions (\ref{chipsi10}) for constant field strength,
we reproduce some terms in the expansions (\ref{chipsi11}).
However, in $\hat \chi$ there is an additional piece linear in $z_1$ which cannot arise in the adiabatic
approximation. In the expansion of $\hat \psi_{\bar 1}$ the term linear in $\bz_1$ is expected because in
the exact solution there is actually a linear term in the constant $\cam$.

\subsection{Evaluating the  fluxes}
\label{ss:eflux}

Before going to the explicit computation of the Yukawa couplings
let us evaluate the size of the expected $U(1)$ fluxes in 
F-theory grand unification schemes. Flux quantization demands 
\beq
\frac {1}{2\pi} \int_{\Sigma \subset S} \!\!\!\! F \  = \  {\tilde m}   \quad ; \quad 
\frac {1}{2\pi} \int_{\Sigma_a}\! F^{(1)}\ =  \  \tilde n^{(1)} \quad
; \quad \frac {1}{2\pi } \int_{\Sigma_b}\! F^{(2)}\ = \  \tilde m^{(2)}
\label{flujillos}
\eeq
where $\tilde m$, $\tilde n^{(1)}$ and $\tilde m^{(2)}$ generically denote flux quanta for the bulk 
and $U(1)_i$ fluxes. On the other hand, the GUT gauge coupling constant is given by 
%
\beq
\frac {1}{\alpha_G}\ = \ M_*^4 \int_S  \omega\wedge \omega \ =\
 {\rm{Vol}}(S) M_*^4 \ =\  R^4 M_*^4 \ .
\label{alphagut}
\eeq
where $R$ is the overall radius of the manifold $S$.
We then estimate  for the fluxes
\beq
F = 2\pi  \sqrt{\alpha_G} M_*^2 {\tilde m} \quad  ; \quad 
F^{(1)} =  2\pi \sqrt{ \alpha_{G}} M_*^2 {\tilde n^{(1)}} \quad ;\quad  
F^{(2)}  =  2\pi  \sqrt{\alpha_{G}} M_*^2 {\tilde m^{(2)}} \ .
\label{flujetes}
\eeq
Here we have assumed that the volume of each matter curve is controlled by 
the overall size $R$, since they are embedded in $S$.
Recall that standard MSSM gauge coupling unification gives
$\alpha_G\simeq 1/24$, for the conventional  gauge group normalization 
$\Tr T^2=1/2$, with generators $T$ in the fundamental of $SU(K)$.
Thus, the compactification scale is only slightly below the F-theory scale $M_*$.

Equipped with the above estimates we can characterize more precisely the parametrization
of the field strengths. For instance, we conclude that the total constant coefficients
are generically given by
\beq
\cam =   2\pi \sqrt{\alpha_G}(q \tilde m  + q_2 \tilde m^{(2)})M_*^2
\quad ;\quad
\cn  =  2\pi \sqrt{\alpha_G} (q \tilde n + q_1 \tilde n^{(1)})M_*^2
\label{estfluxm}
\eeq
where $q$ and $q_i$ are respectively the bulk and $U(1)_i$ charges.
Similarly, for the total linear coefficients we can write
\beq
\alpha_1   =  2\pi \sqrt{\alpha_G}(q \tilde \a_{1B}  +  q_2 \tilde \a_1^{(2)}) \frac {M_*^2}{ R}
\quad ; \quad
\alpha_2   =   2\pi \sqrt{\alpha_G} (q \tilde \a_{2B}  + q_1 \tilde\a_2^{(1)}) \frac {M_*^2}{ R}
\label{estfluxa}
\eeq
where $\tilde \a_{iB}$ and $\tilde\a_i^{(j)}$ are adimensional constants that come respectively 
from bulk and $U(1)_i$ fluxes.

Recall that on the curves $\Sigma_a$ and $\Sigma_b$ the effective wave functions,
in the holomorphic  gauge,  depend only on 
parameters given by  bulk quantities. Specifically they are functions of 
\beqa
\cam_a  =  2\pi \sqrt{\alpha_G}q_a \tilde m M_*^2 \quad & ; & \quad 
\alpha_{1a} \equiv \alpha_a   =  2\pi \sqrt{\alpha_G}q_a \tilde \a_{1B} \frac {M_*^2}{ R} 
\label{fluxsa} \\  
\cn_b  =  2\pi \sqrt{\alpha_G}q_b \tilde n M_*^2 \quad & ; & \quad 
\alpha_{2b} \equiv \alpha_b   =  2\pi \sqrt{\alpha_G}q_b \tilde \a_{2B} \frac {M_*^2}{ R} 
\label{fluxsb}
\eeqa
Other coefficients such as say, $\cam_b  =  2\pi \sqrt{\alpha_G}(q_b \tilde m + e_2  \tilde m^{(2)}) M_*^2$,
do not appear in the wave functions in the holomorphic gauge $\hat{\ca}$. 
On the other hand, the parameters for the curve $\Sigma_c$ depend on bulk and $U(1)_i$ fluxes
according to (see appendix A)
\beqa
\cam_c  =  2\pi \sqrt{\alpha_G}(q_c \tilde m - e_2  \tilde m^{(2)}) M_*^2 \quad & ; & \quad
\cn_c  =  2\pi \sqrt{\alpha_G}(q_c \tilde n - e_1  \tilde n^{(1)}) M_*^2 
\label{fluxsc} \\ 
\alpha_{1c} =  2\pi \sqrt{\alpha_G}(q_c \tilde \a_{1B} -e_2 \tilde \a_1^{(2)})\frac {M_*^2}{ R} 
\quad & ; & \quad 
\alpha_{2c} =  2\pi \sqrt{\alpha_G}(q_c \tilde \a_{2B} -e_1 \tilde \a_2^{(1)})\frac {M_*^2}{ R} 
\nonumber
\eeqa
In the following we will use $\Delta=(\cam_c+\cn_c)/2$ instead of $\cn_c$, and
$\delta=(\a_{1c}+\a_{2c})/2$ in place of $\a_{2c}$, and we
will denote $\a_{1c} \equiv \a_c$. The decomposition of the quadratic and higher order coefficients
of $\cf$ is completely analogous. Observe that gauge invariance imposes constraints such as
$\cam_a + \cam_b + \cam_c =0$.

Note that the bulk and $U(1)_i$ charges, $q$, $q_1$, and $q_2$, depend on the normalization of the gauge 
coupling constants. Consider for example the case of a bulk hypercharge $U(1)_Y$ with 
$q_Y$ integer normalization such that \mbox{$q_Y(Q_L,U,D,L,E)=e_Y(1,-4,2,-3,6)$}. 
Then,  \mbox{$\Tr Q_Y^2 =(12+18)e_Y^2=30e_Y^2$}, evaluating at a $SU(5)$ 5-plet. In order to get 
the standard $SU(5)$ normalization with $\Tr T^2=1/2$, a normalization
factor $e_Y=1/\sqrt{60}\simeq 0.13$ is needed.
The same exercise for $U(1)_{B-L}$ in $SO(10)$ yields a factor $e_{B_L}=1/\sqrt{24}$, 
with assignments \mbox{$q_{B-L}(Q_L,Q_R,L,R)=e_{B-L}(1,-1,-3,3)$}.

The  normalization factors for $q_{1,2}$ are found in an analogous way,  
taking into account the enhanced gauge symmetry at each matter curve. Consider for example matter curves 
at which an $SU(5)$ symmetry is enhanced to $SO(10)$ or $SU(6)$.
This means that there are branchings $SO(10)=SU(5)\times U(1)_{1,2}$
or $SU(6)=SU(5)\times U(1)_{1,2}$. One finds normalization constants
$1/\sqrt{40}$ and $1/\sqrt{60}$ respectively, with matter fields having charges $\pm 1$.
In the case of $SO(10)$ with matter curves enhancing to
$E_6$ or $SO(12)$ one finds $1/\sqrt{20}$ and $1/\sqrt{8}$ respectively.  
These factors must be taken into account in the explicit computation of coupling constants.

There is an additional constraint on the fluxes in the bulk coming from the BPS condition in eq.(\ref{bpscond})
which now reduces to $\omega \wedge F =0$.
In particular, locally this condition implies $M+N = 0$ for constant $F$.  
Nevertheless, in what follows we will not impose this constraint 
so that we can keep track of the effect of all flux parameters.

\begin{figure}
\begin{center}
\includegraphics[scale=0.5]{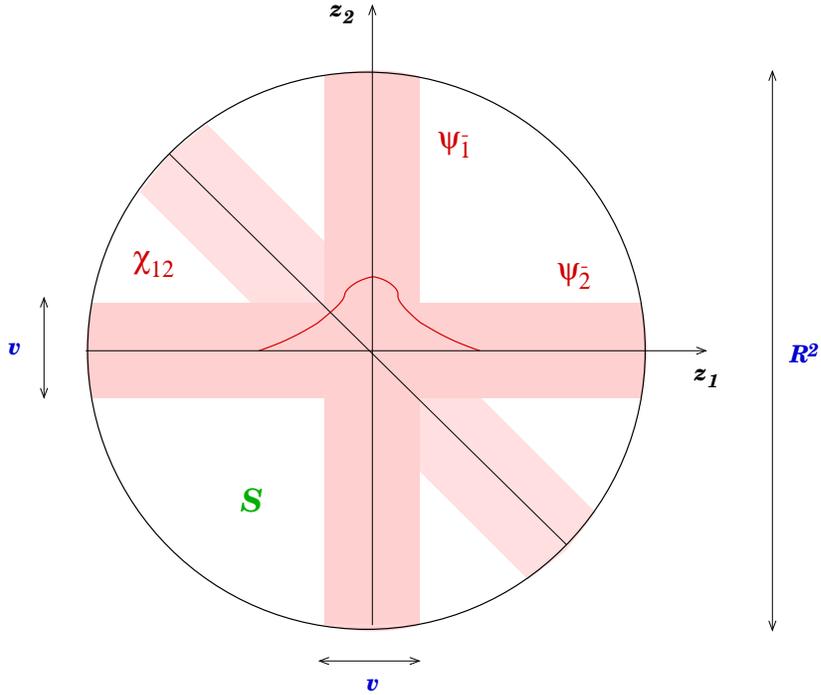} 
\end{center}
\caption{\small Triple intersection of three  matter curves.}
\label{calvicie}
\end{figure}


\section{Yukawa couplings}

\subsection{Computing  Yukawa couplings}
\label{ss:yukint}

We are interested in evaluating the Yukawa coupling of
three chiral fields coming from three intersecting matter curves
locally described by $z_1=0$, $z_2=0$ and $z_1+z_2=0$ in the
surface $S$. The piece of the superpotential relevant for Yukawa couplings will be 
\beq
W_Y   =   M_*^4\int_{ S} \Tr( {\pmb A} \wedge {\pmb A}\wedge {\pmb \Phi})
\label{superpot1}
\eeq
where $M_*$ is the typical mass scale characteristic of the
supergravity limit of F-theory. The Yukawa couplings are
obtained as overlap integrals  over $S$ of the three wave functions 
involved. In principle such a computation requires a 
knowledge of the wave functions over the whole complex surface $S$.
On the other hand, we know that the wave functions are peaked 
around the local curves $z_1=0$, $z_2=0$ and $z_1+z_2=0$ so
that  the coupling is dominated by the 
region around the origin, $z_1=z_2=0$, where the three curves meet. 
If this is correct, a local knowledge of the wave functions
of the type discussed in the previous sections will be 
sufficient to evaluate the Yukawa couplings. We will thus be 
interested on  overlap integrals  of the form\footnote{See however the note added at the end of the paper.}
\beq
Y =  M_*^4\int_S \! d^2z_1 d^2z_2  \, \psi_{\bar 1}\, \psi_{\bar 2}\, \varphi_{12}
\label{yuk1}
\eeq
involving zero modes of the curves $z_1=0$, $z_2=0$, and $z_1+z_2=0$
respectively, $z_i$ being the local coordinates. 
Given this structure it is natural to assign the physical, e.g. quark/lepton, fields to
$\psi_{\bar \imath}$ zero modes and $\varphi_{12}$ to the Higgs boson 
and we will assume this in what follows.

We will then take $\psi_{\bar 1}$, $\psi_{\bar 2}$, and $\varphi_{12}$ to be the zero modes localized
at the curves $\Sigma_a$, $\Sigma_b$ and $\Sigma_c$ respectively.    
By supersymmetry the wave function of $\varphi_{12}$ is equal to that of $\chi_c$.
We have seen in the previous sections that, in the holomorphic  gauge,  the relevant wave functions 
in the presence of variable fluxes take the general form
\beqa
\psi_{\bar 1}  & =&  - f(z_2)e^{-e_1m^2|z_1|^2} \, G_a(z_1,{\bar z}_1; q) \nonumber \\
\psi_{\bar 2}  & =&  - g(z_1)e^{-e_2m^2|z_2|^2} \, G_b(z_2,{\bar z}_2; q) \label{fdo1} \\
\varphi_{12}   & =&    e^{-\frac{em^2}{\sqrt{2} }
|z_1+z_2|^2} \, G_c(z_1, \bz_1, z_2, \bz_2;q,q_1,q_2) 
\nonumber
\label{rayos}
\eeqa
where $G_I$, $I=a,b,c$, are functions which can be computed to any desired order 
both in fluxes and derivatives of fluxes. Recall that $G_a$ and $G_b$ depend only on bulk 
charges whereas $G_c$ depends on all charges. In absence of fluxes one simply has $G_a=G_b=G_c=1$.  
Here $f(z_2)$ and $g(z_1)$ are holomorphic functions. 
As in \cite{hv3} we will choose a basis in which they are given by
$f_k=(z_2/R)^{3-k}$ and $g_\ell=(z_1/R)^{3-\ell}$, $k,\ell=1,2,3$, corresponding to the three 
generations of quarks and leptons. Since there is only one  Higgs field
we can take the corresponding holomorphic function to be a constant.
We then have to perform the integral
\beq
Y_{k\ell}  = M_*^4\int_S \! d^2z_1 d^2z_2 \, e^{-em^2 (|z_1|^2+ |z_2|^2+
\frac {1}{\sqrt{2}}|z_1+z_2|^2)}\, f_k(z_2)\, g_{\ell}(z_1)\, 
 G(z_1,\bz_1, z_2, \bz_2)  \ ,
\label{yuk2}
\eeq
where $G=G_a G_b G_c$. To simplify the analysis we have set $e_1=e_2=e$.

An important point to remark is that  turning on  fluxes does not
induce mixing in the wave functions among different flavors. Indeed, as
seen e.g.  in eq.(4.3),  the flux corrections in $G_a$($G_b$) 
do not introduce additional holomorphic dependence on $z_2$($z_1$) which would 
signal generation mixing in the wave functions. This is an important
simplification because otherwise we would need an additional diagonalization of
wave functions in order to extract the physical couplings from 
eq.(\ref{yuk1}).

The measure in the Yukawa integral can be thought to be
\beq
d\mu=   d^2z_1 d^2z_2 \, e^{-em^2(|z_1|^2 + |z_2|^2 +
\frac{1}{\sqrt2} |z_1+z_2|^2 )}
\label{measure1}
\eeq
Clearly, the third exponential, due to the zero mode on $\Sigma_c$, 
breaks the symmetry under separate $U(1)$ rotations  $z_i\rightarrow e^{i\theta_i}z_i$.
Instead, there is only invariance under the diagonal $U(1)$. This is enough to show 
that without non-constant fluxes, in which case $G$ cannot depend separately on the antiholomorphic variables, 
the only non-vanishing Yukawa coupling is $Y_{33}$ because $f_3=g_3=1$. 
Thus, the heaviest third generation of quarks and leptons will acquire masses through $Y_{33}$. 

As pointed out originally in \cite{hv3}, to generate non-vanishing Yukawa couplings
for all families it is necessary to turn on non-constant background fluxes.
To see this it is useful to rewrite the measure as
\beq
d\mu=  \frac14  \, d^2u \, d^2w \, e^{-em^2(\frac12|u|^2 +\frac{1}{2s}|w|^2)}
\label{measure2}
\eeq
where $s={\sqrt2}-1$. As before, $w=z_1+z_2$ and $u=z_1 - z_2$.
In presence of variable fluxes the function $G$ can furnish adequate powers of $\bar w$ and $\bar u$ so that
the integrand becomes invariant under separate phase rotations of $w$ and $u$. The couplings $Y_{k\ell}$
will thus be non-zero and the light generations will gain masses and mixings.
We have introduced the parameter $s$ in order to study also the case $s=1$,
which corresponds to ignoring the zero mode exponential from 
$\Sigma_c$. In this situation the measure becomes invariant under separate $U(1)$ rotations $z_i\rightarrow e^{i\theta_i}z_i$
and there will be additional cancellations when computing the integrals.

In performing the integration we will assume that the 
width of the matter curves is determined by the F-theory scale $M_*$, this means $v=1/(em^2)=1/M_*^2$.
Consistency of the local analysis requires the matter curves to be well localized within $S$.
This amounts to the condition $v/R^2 \ll 1$, which is approximately valid
for $v=1/M_*^2$ because $1/R^2 = \sqrt{\alpha_G} M_*^2$, and $\sqrt{\alpha_G} \simeq 0.2$.    
In practice we will evaluate the integrals over $S$, with the above measure $d\mu$, by extending
$|w|$ and $|u|$ to infinite radius. The main contribution to the integrand still comes from the region
near the origin because the measure is sufficiently peaked. The upshot is that in the end all integrals
can be done exactly.  

Without varying fluxes there is only one non-vanishing Yukawa coupling $Y_{33}$ for the third generation
which may be explicitly estimated as
\beq
Y_{33}^{(0)} = M_*^4\int_S \! d^2z_1 d^2z_2 \, e^{-M_*^2(|z_1|^2+|z_2|^2+\frac {1}{\sqrt{2}}|z_1+z_2|^2)}  
\,  =    \, \frac {\pi^2}{(1+\sqrt{2})}  \ .
\label{yuk3}
\eeq
To get the physical Yukawa coupling we really need to work 
with wave functions normalized to unity, but to actually normalize our wave functions we would need a global 
knowledge of them over all $S$. We  can however make an estimate by neglecting the effect 
of fluxes and computing the norm of the $\psi_{\bar \imath}$ wave functions from 
\beq
\cc \simeq   M_*^4\int_S \,  e^{-2M_*^2|z_i|^2} \, =  \, \frac{\pi}{2}M_*^2R^2
\label{normfdo}
\eeq
Thus, the normalized $\psi_{\bar \imath}$ wave functions are
obtained multiplying our wave functions by the normalization factor $\cc^{-1/2}$.
Similarly computed, the normalization for $\chi_{12}$, arising in the curve $\Sigma_c$, is found to be
$\cc^\prime=\cc/\sqrt2$. We then obtain the normalized third generation Yukawa coupling 
\beq
Y_{33}^{norm}  \, \simeq \,   
\frac{2^{1/4} \pi^2}{(1+\sqrt{2})}
\left(\frac{2 }{\pi M_*^2  R^2}\right)^{\!\!\! 3/2} 
\ =\ \frac {2^{7/4} \pi^{1/2}}{(1+\sqrt{2})} \alpha_G^{3/4}
\ \simeq \ 0.23
\label{yuktop}
\eeq
where we have taken the $SU(5)$ value $\alpha_G=1/24$.
The  $\alpha_G^{3/4}$ dependence in eq.(\ref{yuktop}) was previously noted in \cite{bhv2}.

The $Y_{33}$ Yukawa just computed is given at the unification scale.
Taking into account QCD renormalization effects down to the weak scale there
is an extra factor  $\simeq 3$ in the case of quarks  so that one obtains
\beq 
m_t\ \simeq  \ 0.23 \times 3 \times \langle H_u \rangle \ = \ 
0.69 \times 170 \ \sin\beta  \ \simeq \ 117 \ {\rm GeV}
\label{eltop}
\eeq
where in the last step we have assumed a large value for $\tan\beta=\langle H_u/H_d \rangle$.
This is in  reasonable  agreement with experiment,
given the uncertainties. A large value for $\tan\beta $ is required to understand 
within this scheme the relative smallness of the masses of b-quark and 
$\tau$ lepton compared to the top. For them one finds
\beq 
m_b\ \simeq \ \frac{m_t}{\tan\beta} \   \quad ;\quad 
m_{\tau } \ \simeq \ \frac{m_t}{3\tan\beta} 
\label{bytau}
\eeq
which  gives reasonable  agreement for $\tan\beta \simeq 35-45 $.
Note that the tau lepton is lighter by a factor $\simeq 3$ due to the absence of QCD renormalization.

There are also subleading contributions to $Y_{33}$ from flux corrections which
appear even for constant flux (see appendix B.1). We will eventually neglect
all subleading corrections so for consistency we will only keep the leading term in $Y_{33}$. 
When we calculate the rest of the Yukawa couplings we will then normalize them
relative to the 3rd generation Yukawa in eq.(\ref{yuk3}).

\subsection{The case of a constant $\chi_c$ wave function}
\label{ss:chic}

We study first this simple case because it has some interesting features by itself.
Furthermore, a constant wave function is unlocalized and hence 
could serve to give an idea of the results to be expected for
Yukawa couplings in which the third particle,
presumably the Higgs field, lives in the bulk rather than in a 
localized matter curve. Such type of couplings do appear in 
type IIB and F-theory models in which the base $S$ is not del Pezzo.  

When $\chi_c$ is a constant, taken equal to one, the Yukawa couplings are determined by
\beq
Y_{k\ell}\left|_{\chi_c=1}\right. 
= M_*^4\int_S \! d^2z_1 d^2z_2 \, e^{-em^2(|z_1|^2+ |z_2|^2)} \, f_k(z_2)\, 
g_{\ell}(z_1)\,
 G_a(z_1,\bz_1;q)G_b(z_2, \bz_2; q) 
\label{yuknoc}
\eeq
where $q$ denotes the bulk charges. Substituting the expressions for $G_a$ and $G_b$, which may be extracted 
from the wave functions in appendix \ref{a:vflux}, leads to
\beq
Y_{k\ell}\left|_{\chi_c=1}\right.  \ = \  \pi^2 \, \delta_{k-3,\ell -3} 
\label{cancelara}
\eeq
Hence, the flux-induced distortion of the wave functions does not give rise 
to any new couplings, only the coupling $Y_{33}$ which is there already 
for constant fluxes is non-vanishing. 
This is true for any order in the flux expansion. In the next section and in appendix \ref{b:yukresu} 
we will provide some examples of the cancellation in the expansion of the $Y_{k\ell}$.
The result can also be proven analytically. In fact, notice that in (\ref{yuknoc}) the integrals in $z_1$
and $z_2$ decouple so that it suffices to show that 
$I_\ell=\int \! d^2z_1 e^{-em^2|z_1|^2} z_1^{3-\ell} G_a$ vanishes when $\ell=1,2$. The key point is that
$ e^{-em^2|z_1|^2} G_a$ can be written as $\frac{1}{z_1}\bar\partial_{\bar 1}F_a$, as explained in appendix A.2.
The function $F_a$ can be extracted explicitly, in particular it
goes to zero when $|z_1| \to \infty$ and to 1 when $|z_1| \to 0$. 
It is then easy to show that $I_\ell=0$ for $\ell=1,2$.

The main conclusion is that in order to get
non-trivial fermion mass hierarchies one needs  all three wave functions 
to be localized on matter curves. We then proceed to this most 
interesting case of three overlapping localized wave functions.

\subsection{ Yukawa matrices}
\label{ss:yukmat}

The physical Yukawas are obtained evaluating the overlapping integral in eq.(\ref{yuk2})  
which is dominated by the region close to the intersection point.
The heavy task is to compute the function $G$  by substituting the wave functions found
in the previous sections expanded in powers both of the flux and derivatives of the flux.
In the end each Yukawa coupling reduces to a sum of Gaussian integrals that can evaluated
analytically. As expected, $Y_{ij}$ and $Y_{ji}$ are related by an appropriate exchange of
flux parameters.

To begin we have considered the simplest case in which the field strengths are expanded
only to linear order. This means that we only take into account the 
first derivative of the fluxes (i.e. the $\alpha_I$ and $\delta$ parameters) 
and neglect the effect of higher derivatives. 
The integrals can be determined exactly. For example, the coupling $Y_{23}$ is found to be
\beq
Y_{23} = \frac{v^2}{3R}\big[\, (s-1)^2 \bar\a_a + (s^2-1) \bar\a_b -3s \bar\a_c + s(2s+3)\bar\d\, \big]
\label{y23exact}
\eeq
where $s=\sqrt2-1$ is the parameter appearing in the measure (\ref{measure2}). 
This coupling is normalized with respect to $Y_{33}^{(0)}=\pi^2 s$.
This exact expression also shows that when $s=1$ the terms that depend purely on
$\a_a$ and $\a_b$ completely drop out. In all couplings it happens that for $s=1$ 
all pieces involving only parameters of the curves $\Sigma_a$ and $\Sigma_b$ do cancel out.   
This implies that when $\chi_c=1$, the only coupling that survives is $Y_{33}$.

In appendix \ref{b:f1} we display the leading terms in the expansion in $\a$-fluxes
for each entry of the Yukawa matrix, normalized with respect to $Y_{33}^{(0)}$.
Some of the elements $Y_{ij}$ have complicated expressions in terms of the
flux parameters but the pattern behind can be easily understood.
Schematically, the couplings turn out to be
\beq
Y_{ij} \sim \left(\frac{v^2\bar\a}{R}\right)^{3-i} \left(\frac{v^2\bar\a}{R}\right)^{3-j}
\label{yijalpha}
\eeq
As explained in section \ref{ss:eflux}, we have for instance $\a_a=2\pi\sqrt{\a_G}q_a \tilde\a_{1B} M_*^2/R$.
Therefore, we find  $v^2\a_a/R=2\pi \a_G q_a \tilde \a_{1B}$, because $v=1/M_*^2$ and
$M_*^2 R^2 = 1/\sqrt{\a_G}$. 

More generally, the corrections to the Yukawa couplings due to first derivatives of the fluxes have the general form 
\beq
Y_{ij} \ \simeq \ \xi_{ij}\, \left(2\pi\alpha_G(aq + a'q^\prime)\right)^{(3-i)}
\left(2\pi\alpha_G(aq+a'q^\prime)\right)^{(3-j)}
\label{yukflux1}
\eeq
Here we have simply replaced the $\tilde \a_{iB}$ and the $\tilde \a_i^{(j)}$ of section \ref{ss:eflux}
by generic constants $a$ and $a^\prime$ in order to get an idea of the structure. 
The $\xi_{ij}$ are numerical coefficients appearing upon integration
which are typically in the range $0.1-10$, as may be seen in appendix \ref{b:f1}.
The constants $q$ and $q^\prime$ are the bulk and matter curve $U(1)$ charges respectively.
Recall that for the fields in matter curves $\Sigma_a$ and $\Sigma_b$, which include 
quarks and leptons,  one has $q^\prime=0$ and the corresponding $\alpha_a$ and $\alpha_b$ 
parameters only depend on the bulk $U(1)$ charges. This is not the 
case for the matter curve $\Sigma_c$, the parameters $\alpha_c$ and $\delta$ 
do depend on the matter curve charges. Note that the normalization of the 
$U(1)$ charges is relevant here. As we explained, for the $SU(5)$ case 
 for integer  hypercharge  there is a normalization factor $1/\sqrt{60}$ and the 
$U(1)$'s on the matter curves containing $10$'s and ${\bar{5}}$'s
have normalization $1/\sqrt{40}$ and $1/\sqrt{60}$ respectively.

We have just discussed the general form of each of the terms in the Yukawa couplings shown
in appendix \ref{b:f1}. To get more accurate results we would need to specify the
different flux parameters for the three matter curves involved. In particular, we would need a
precise knowledge of how the $U(1)$ field strengths vary in the vicinity of the
intersecting points. In principle, given a set of assumptions about the
derivatives of fluxes on the different matter curves in a concrete 
model, the formulas in appendix B will allow us to compute the different Yukawa couplings. 

It is already quite encouraging that a hierarchical structure of fermion masses seems to be built in.
Using  eq.(\ref{yukflux1}) we can further estimate the Yukawa couplings by taking into account the
normalization of the $U(1)$ charges explained in section \ref{ss:eflux}. To this end we will write
the bulk charges as $q=\hat q e_B$, where $\hat q$ is an integer and $e_B$ is the  bulk charge
normalization. We will similarly write $q^\prime=\hat q^\prime e$, where $e$ is the normalization of the
appropriate $U(1)$ on the matter curve, and reabsorb the ratio $e/e_B$ into the coefficient $a^\prime$.  
It is also convenient to introduce $\alpha_{U(1)}= e_B^2 \alpha_G$, which corresponds to the $U(1)$ 
fine structure constant normalized for integer charges of massless fields. We then conclude that the
Yukawa matrix has the form 
\beq
Y^{(1)} \ \sim \
\left(
\begin{array}{ccc}
 \e^4\eta^4  &  \e^3\eta^3   &  \e^2\eta^2 \\
 \e^3\eta^3 & \e^2\eta^2  &  \e\eta \\
\e^2\eta^2 & \e \eta  &  1
\end{array}
\right) \quad ;\quad \e=2\pi\sqrt{\alpha_{U(1)}} \quad ;\quad
\eta= \sqrt{\alpha_G}
\label{flx2}
\eeq
In the terminology of \cite{hv3}, 
we can say that the physical parameter $\e$, which is tied to the $\a_I$ coefficients, 
controls the flux expansion. The parameter $\eta$ is related to powers of the width $v$
and the overall radius $R$ that appear in the couplings and controls instead the derivative
expansion. Taking $\alpha_G=1/24$ and $e_B=e_Y=1/\sqrt{60}$ gives $\e = 0.16$, $\eta =0.20$
so that $\e\eta \simeq 1/31$. We then find
\beq
Y_{ij} \ \simeq \ \xi_{ij}\, \left( \frac {a\hat q+a'\hat q^\prime}{31}\right)^{(3-i)}
\left( \frac {a\hat q+a'\hat q^\prime} {31} \right)^{(3-j)}
\label{yukflux2}
\eeq
Therefore, the fermion hierarchies are roughly of the type 
\beq
(m_3:m_2:m_1) \ \simeq \ (1:10^{-3}(a\hat q+a'\hat q^\prime)^2 : 10^{-6}(a\hat q+a'\hat q^\prime)^4)
\label{jerarquias0}
\eeq
in qualitative agreement with
the observed spectra of quarks and leptons.  In the next chapter
we discuss in slightly more detail to what extent 
this structure may be successful in describing the pattern of quark and lepton masses.

Let us now see what happens if further terms in the derivative expansion 
of the fluxes were non-negligible. In particular, we have studied the corrections to the
Yukawa couplings when the second derivative flux parameters $\beta_I$ and $\rho$ are non-zero.
We found that the couplings $Y_{13}$ and $Y_{22}$ receive leading contributions linear in $\b_I$ or $\rho$.
They also have quadratic corrections, proportional to the constant coefficients $\cam$ and $\cn$ of the various
curves times the $\b_I$ or $\rho$, that are subleading and can be neglected. The leading linear terms are  
\beqa
\hspace*{-5mm}Y_{13} & \! = \!& \frac{v^3}{8R^2}\big[\, 3(s-1)^3 \bar\b_a + 3(s+1)(s^2-1) \bar\b_b -16s^2 \bar\b_c + 
2s(3s^2+8s+6)\bar\rho\, \big]
\label{y13exact} \\[2mm]
\hspace*{-5mm}Y_{22} & \! = \! & \frac{v^3}{8R^2}\big[\, 3(s-1)(s^2-1) (\bar\b_a + \bar\b_b) + 2s(s^2-2)\bar\rho\, \big]
\label{y22exact}
\eeqa
where $s=\sqrt2-1$ as before. Here we notice again that when $\chi_c=1$ the couplings will vanish identically
because in this case $s=1$ while $\b_c=0$ and  $\rho=0$.
The couplings $Y_{12}$ and $Y_{11}$ have leading corrections typically proportional to $\a_I \b_J$ and $\b_I \b_J$ respectively, but
the exact expressions are too long to display.
In appendix \ref{b:f2} we show the numeric results for the extra leading contributions 

To figure out the size of the corrections due to the second derivative flux parameters we will estimate 
them for the case of the $Y_{11}$ and $Y_{22}$ Yukawa couplings. We have
\beqa
Y_{22} & \ \sim \ &
\frac {v^3\bar \beta}{R^2} \simeq \ \ \frac {2\pi v^3 \sqrt{\alpha_{U(1)}} (b\hat q+b'\hat q^\prime)M_*^2}{R^4}
= 2\pi \sqrt{\alpha_{U(1)}}\alpha_G  (b\hat q+b'\hat q^\prime) \\[2mm] 
Y_{11} & \ \sim \ &
\frac {v^6\bar \beta^2}{R^4} \simeq \ \  \frac { (2\pi)^2 v^6 \alpha_{U(1)}  (b\hat q+b'\hat q^\prime)^2 M_*^4}{R^8}
= (2\pi)^2\alpha_G^2\alpha_{U(1)} (b\hat q+b'\hat q^\prime)^2 
\label{yukder2}
\eeqa
These terms would contribute to the hierarchy of fermion masses as
\beq
(m_3:m_2:m_1) \ \simeq \ (1:(2\pi)\alpha_G\sqrt{\alpha_{U(1)}}: (2\pi)^2\alpha_G^2\alpha_{U(1)}) \ .
\label{jerarquias}
\eeq
We can evaluate the remaining couplings in the same way (the first order in fluxes $Y_{23}$ and $Y_{32}$ as well). 
Including the zeroth order $Y_{33}$ we obtain the structure
\beq
Y^{(2)} \ \sim \
\left(
\begin{array}{ccc}
 \e^2\eta^4  &  \e^2\eta^3    &  \e \eta^2 \\
 \e^2 \eta^3 & \e \eta^2   &  \e \eta \\
\e \eta^2  &  \e \eta  &  1
\end{array}
\right) 
\label{der22}
\eeq
Since $\e \simeq \eta $, there are hierarchies
$(1: \e^{3}(b\hat q+b'\hat q^\prime):\e^{6}(b\hat q+b'\hat q^\prime)^2)$. We see that for coefficients of order one, 
these corrections will generically dominate over the corresponding terms in the 
flux expansion with only first derivatives of fluxes. 

We can go one step beyond and consider also the effect of terms of order three and
four in the derivative expansion of the fluxes. In this case, to leading order 
there appear contributions to $Y_{12}$, $Y_{21}$ and $Y_{11}$. The results are
given in \ref{b:f34} where we also explain the notation. The additional corrections 
may be approximated by 
\beqa
Y_{11}  &  \sim  &  
\frac {v^5 \bar D}{R^4}  \simeq  \frac {2\pi v^5 \sqrt{\alpha_{U(1)}} (d\hat q+d'\hat q^\prime) M_*^2}{R^8}
= 2\pi\sqrt{\alpha_{U(1)}}\alpha_G^2 (d\hat q+d'\hat q^\prime) 
\label{yukder411}\\[2mm]
Y_{12} &  \sim & Y_{21} \sim  \frac{v^4 \bar C}{R^3}
 =  \frac {2\pi v^4 \sqrt{\alpha_{U(1)}}(c\hat q+c'\hat q^\prime)M_*^2}{R^6}
= 2\pi \sqrt{\alpha_{U(1)}}\alpha_G^{3/2} (c\hat q+c'\hat q^\prime) 
\label{yukder4}
\eeqa
In this case there are new contributions to the Yukawa couplings of the form
\beq
Y^{(3,4)} \ \sim \
\left(
\begin{array}{ccc}
 \e\eta^4  &  \e \eta^3    &   0  \\
 \e \eta^3  &   0   &  0 \\
     0  &  0  &  0
\end{array}
\right) 
\label{der44}
\eeq
Thus, the first generation Yukawa has corrections of order $Y_{11}\simeq \e \eta^4$.

The contributions leading to the terms captured by $Y^{(1)}$ are an explicit evaluation of the
flux expansion of the authors of \cite{hv3}. On the other hand, their derivative expansion 
would correspond to taking the terms linear in $\e$ in $Y^{(1)}$, $Y^{(2)}$ and
$Y^{(3,4)}$ above. Thus, this derivative expansion has the structure
\beq
Y^{DER} \ \sim \
\left(
\begin{array}{ccc}
 \e\eta^4  &  \e \eta^3    &   \e \eta^2 \\
 \e \eta^3  &   \e \eta^2   &  \e \eta \\
     \e \eta^2  &  \e \eta  &  1
\end{array}
\right) 
\label{DER}
\eeq
Note however  that, for instance in $Y^{(2)}$, we also find corrections which do not correspond
to either of both expansions.

As a general conclusion, one observes that, for a given Yukawa matrix element,
the correction due to a higher order term in the derivative expansion will always
dominate over the flux expansion.

\section{Fermion Yukawa couplings in F-theory GUT's}

\begin{figure}
\begin{center}
\includegraphics[scale=0.4]{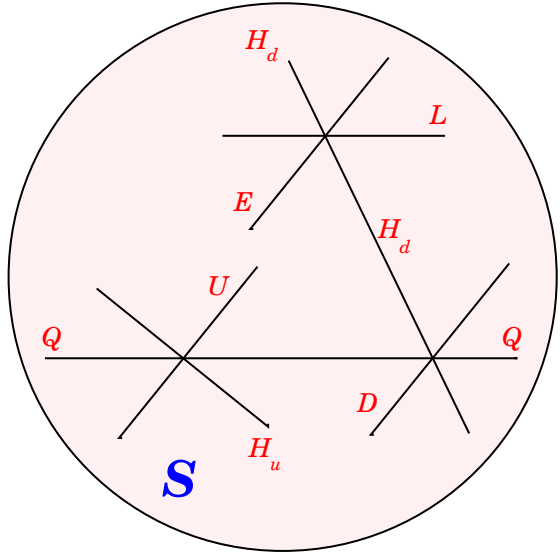}
\end{center}
\caption{\small Intersecting matter curves and Yukawa couplings.}
\label{losacoplos}
\end{figure}

In the previous sections we have studied the zero modes of the
8-dimensional quasi-topological field theory as well as the
computation of the Yukawa couplings using these zero modes,
without specifying any particular geometry nor identifying the
nature of the three particles involved in the couplings.
We did neither specify the bulk $U(1)$ to be considered,
which could be hypercharge in $SU(5)$, \mbox{$B$-$L$} or other in $SO(10)$.
In a  GUT model, the  quark, lepton and Higgs superfields
will be localized in matter curves like those we have described
(see figure \ref{losacoplos}). We will have Yukawa couplings
from a superpotential of the form
\beq
W_{Yuk}\ =\ Y^U_{ij}Q^iU^jH_u\ +\ Y^D_{ij}Q^iD^jH_d\ +\
Y^L_{ij} L^iE^jH_d
\label{supyuk}
\eeq
in an obvious notation. In principle the intersection points
of the different matter curves will be different and,
correspondingly the flux parameters $\alpha_I$,
$\beta_I$, etc., will also be different at each intersection.
The geometry of each given model may constrain the possibilities though.
For example, in an $SU(5)$ GUT the left-handed leptons $L^i$ and
the right-handed $D$-quarks $D^j$ live in the same matter curve.
In other settings that need not be the case. For example, in
a flipped SU(5) setting,  $D$, $U$ and $L$ masses come from independent
couplings $10\times 10\times 5_H$, $10\times {\bar 5}\times {\bar 5}_H$
and $1\times {\bar 5}\times { 5}_H$.

Another point to emphasize is that in previous chapters we have made use
of the possibility of choosing a local basis of holomorphic wave functions
of the canonical form $1,z_i,z_i^2$ at the intersection point. Note however that
in the case of quarks we cannot make use of the freedom to
choose that basis both at the intersection point leading to $U$-quark
masses and that giving rise to $D$-quark masses. On the other hand, if
the holomorphic basis at both points are very different, one
expects very large  CKM mixing angles. This may be an indication
that both points must be quite close in $S$ in order for the
basis to be aligned to give reduced mixing, as required
phenomenologically \cite{hv3}. That points towards further unification into
at least $E_7$ at the F-theory level.

\begin{figure}
\begin{center}
\includegraphics[scale=0.5]{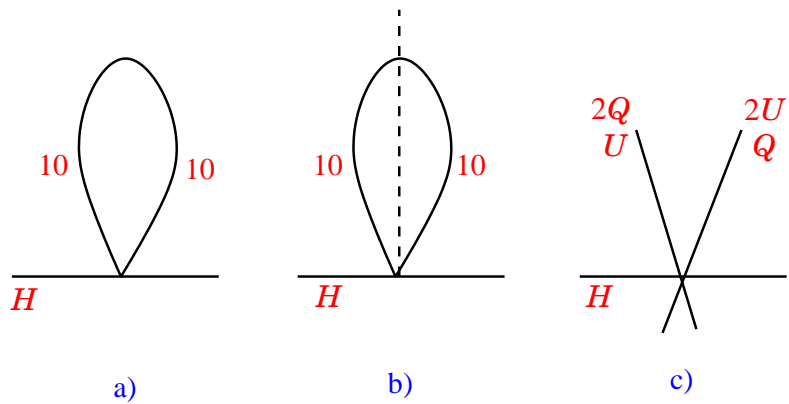}
\end{center}
\caption{\small Matter curves intersecting to provide $U$-quark Yukawas.}
\label{lastres}
\end{figure}

An important issue in F-theory grand unification is the generation
of appropriate Yukawa couplings for the $U$-quarks. After all, one of the
main motivations for going to F-theory GUT's instead of perturbative
IIB orientifolds is that in the former case these couplings are allowed,
while they are perturbatively forbidden in type IIB orientifolds.
In  $SU(5)$ the $U$-quark Yukawas come from
couplings $10^i\times 10^j\times 5_H$, and if such coupling comes from three
distinct matter curves, there can be no diagonal $U$-quark couplings.
This implies that the trace of $Y^U$ vanishes which makes impossible
a hierarchy of $U$-quark masses. In \cite{bhv1,bhv2,hv3}  it was suggested that
the matter curve associated to the $10$'s in $SU(5)$ could
{\it self-pinch} as in figure \mbox{\ref{lastres}-a}, allowing for diagonal entries.
It was noted though \cite{tatar2} that in such configuration the two branches of the
wave functions of the $10$ are independent so that there would be
two independent rank one contributions to
$Y^U$. This would then lead to a rank
two Yukawa matrix, with no automatic hierarchical structure.
In \cite{tatar2} (see also \cite{hv3}) it was suggested that in fact the two
independent branches of the wave functions could be identified by
some symmetry in the geometry (figure \mbox{\ref{lastres}-b}). In such a case the two  rank one
contributions would be identical and rank one (before the addition of flux effects).
It was also argued that these symmetries are ubiquitous in F-theory and
correspond to non-trivial monodromies. Another  alternative in order to obtain
diagonal entries in $U$-quark Yukawa couplings was also suggested in  \cite{fi}.
It is more easily described in $SO(10)$ but it also applies to $SU(5)$.
In the context of $SO(10)$ the Yukawa couplings come from terms $16^i\times 16^j\times 10_H$.
If one associates both $16$'s in the coupling to two matter curves $\Sigma_a$ and $\Sigma_b$,
and one allows for appropriate \mbox{$B$-$L$} flux with opposite restriction on the curves,
the massless spectrum splits as in figure \mbox{\ref{lastres}-c}. One curve has matter content $2Q + U$, and the
other $2U+Q$, and this splitting allows for diagonal couplings. In fact, as already noted in
\cite{fi}, both matter curves could be local branches of some self-pinched
matter curve, as in figures \mbox{\ref{lastres}-a} and \mbox{\ref{lastres}-b}.
An interesting feature of this possibility is that the mixing of the first generation 
with the other two is expected to be suppressed. 
In what follows we will not specify the particular scheme for the understanding of the $U$-quark Yukawa couplings. We will 
just consider that the approximate rank one structure already assumed in the previous sections does apply.

In this section we make a preliminary analysis of the
application of our previous results to the description of quark
and lepton spectra. We would like to see
to what extent flux distortion may explain the data.
 Let us first consider for definiteness the case of
a $SU(5)$ GUT broken down to the SM by fluxes along the hypercharge
direction. Let us first see whether the flux-induced  distortion 
of wave functions due to first derivatives of fluxes is enough 
to describe the observed structure of quarks and leptons. 
In order to get manageable results  we will  first  assume for simplicity
that the fluxes going through the third matter curve 
are approximately constant, i.e. $\alpha_c=\beta_c=\delta=\rho=0$
(we will also denote the subscripts $(a,b)$ as $(1,2)$ hereafter).
Under these circumstances the formulas in appendix \ref{b:f1}
substantially simplify. In particular, the diagonal entries reduce to
\beqa
Y_{22} & \simeq   &
\e^2\eta^2\left( 0.067(Y_R^2{\bar a}_1^2+Y_L^2{\bar a}_2^2)+0.11\ Y_RY_L{\bar a}_1{\bar a}_2\right) 
\label{fluxsm1} \\[2mm]
Y_{11} &  \simeq  &
\e^4\eta^4\left( 0.10(Y_R^4{\bar a}_1^4+Y_L^4{\bar a}_2^4)+0.09Y_RY_L{\bar a}_1{\bar a}_2(Y_R^2
{\bar a}_1^2+Y_L^2{\bar a}_2^2)+0.12Y_L^2Y_R^2{\bar a}_1^2{\bar a}_2^2  \right)
\nonumber
\eeqa
where $Y_R,Y_L$ are the (integer) hypercharges of right-handed fermions 
(on matter curve $\Sigma_a$) and left-handed ones (on matter curve $\Sigma_b$).
Recall that
\beq
\e \ =\ 2\pi \sqrt{\alpha_Y}\ = \frac {2\pi}{\sqrt{60}}\sqrt{ \alpha_G }
\   \simeq \ 0.16   \quad  ; \quad \eta \  = \  \sqrt{\alpha_G} \ \simeq \ 0.20
\eeq
where we have taken into account the hypercharge normalization factor $e_Y=1/ \sqrt{60}$. 
Note that, as we mentioned before, the wave functions in matter curves $\Sigma_a$ and $\Sigma_b$
in the holomorphic gauge 
have no dependence on the $U(1)_i$ fluxes, they only depend on the bulk fluxes
which go in this case along hypercharge. Here $a_{1,2}$ are the adimensional
constants parametrizing the variation of the bulk flux close to the intersection 
point. Note that in principle these parameters may be different for the 
three different Yukawa couplings, i.e. there are 
$a^{U,D,L}_{1,2}$.  To get an idea of the size of the Yukawas let
us for the moment assume that $a^{U.D,L}_1\simeq a^{U,D,L}_2\simeq a^{U,D,L}$.
Note that in this case that we neglect the flux variation for the
Higgs matter curve, the Yukawa matrix is strongly dependent on the
hypercharge of the quarks and leptons involved in the couplings.
Since the maximum value of the quark and lepton hypercharges is
$|Y_{max}|=6,4,2$ respectively for leptons and $U$ and $D$-quarks, 
one expects larger effects for leptons, $U$-quarks and $D$-quarks 
in that order.

Inserting the  values of the hypercharges 
for  the different particles involved in each Yukawa coupling leads to 
the results in  table \ref{jerarsm1} for the diagonal Yukawas.
\begin{table}[htb] \footnotesize
\renewcommand{\arraystretch}{1.25}
\begin{center}
\begin{tabular}{|c||c|c|c|}
\hline  Yukawa    &   $Y_{33}$  &  $Y_{22}$  &   $Y_{11}$  \\
\hline\hline
 $ Y^U $ &
   1  & $0.7 (a^U)^2 \times 10^{-3}$   &  $2.2  (a^U)^4\times 10^{-5}$  \\
\hline
  $Y^U$(exp) &   1   & $(3-4)\times 10^{-3}$     &   $(0.5-1.6)\times 10^{-5}$    \\
\hline\hline
 $Y^L$   &   1   &  $1.1 (a^L)^2 \times 10^{-3}$   &   $1.1 (a^L)^4 \times 10^{-4}$ \\
\hline
 $Y^L$(exp) &  1  &   $5.9\times 10^{-2}$   & $2.8\times 10^{-4}$    \\
\hline\hline
 $Y^D$   &    1   &   $0.6   (a^D)^2 \times 10^{-3}$   &    $ 3.2  (a^D)^4\times 10^{-6}$ \\
\hline
$Y^D$ (exp) & 1 &    $(1-3)\times 10^{-2}$      &   $(0.6-1.8)\times 10^{-3} $ \\
\hline
\end{tabular}
\end{center}
 \caption{\small Hierarchies of fermion masses from the flux expansion.
First order in derivatives.}
\label{jerarsm1}
\end{table}
The couplings are normalized to the one of the corresponding third generation particle. We also
show for comparison experimental results for that hierarchy evaluated at the
electroweak scale from \cite{fritzsch}. The results for the $U$-quarks 
hierarchies are encouraging, for values $a^U,a^L \simeq 1$ one can describe reasonably  well
the observed  pattern. For the case of charged leptons the mass of the electron is again well described
for $a^L\simeq 1$. However, the mass of the muon would turn out too
light unless $a^L\simeq 7.3$, which would be quite large and incompatible with the 
$a^L\simeq 1$ required for the electron. Thus, the correct numerical description 
would require some further contribution for the muon. Alternatively, it could be that for
charged leptons neglecting the flux variation coming from the Higgs matter curve is not the
correct assumption. 
For the case of the  the $D$-quark hierarchies one would need  slightly large values $a^D\simeq 5.8$
and $a^D\simeq 4$ for $Y_{22}$ and $Y_{11}$ respectively.

Let us now  explore what would be the effect of higher order terms in the 
derivatives of the fluxes. If we consider second order in 
derivatives there are extra corrections which may be extracted from
appendix \ref{b:f2}. We will again set to zero all flux parameters from the curve
$\Sigma_c$. The corresponding diagonal terms are found to be
\beqa
Y_{22} &  \simeq  &
\e\eta^2\ 0.18\left( Y_R{\bar b}_1+ Y_L{\bar b}_2\right) \label{fluxsm2} \\[2mm]
Y_{11} &  \simeq  &
\e^2\eta^4  \   (-0.23 ( Y_R^2 {\bar b}_1^2+Y_L^2{\bar b}_2^2)
+0.29 Y_LY_R{\bar b}_1{\bar b}_2 ) 
\nonumber
\eeqa
Again taking $b^{U,D,L}_1\simeq b^{U,D,L}_2\simeq b^{U,D,L}$ yields contributions as in table \ref{jerarsm2}.
\begin{table}[htb] \footnotesize
\renewcommand{\arraystretch}{1.25}
\begin{center}
\begin{tabular}{|c||c|c|c|}
\hline  Yukawa    &   $Y_{33}$  &  $Y_{22}$  &   $Y_{11}$  \\
\hline\hline
 $ Y^U $ &
   1  & $ 3.4 (b^U)\times 10^{-3}$   &  $2.1 (b^U)^2 \times 10^{-4}$  \\
\hline\hline
 $Y^L$   &   1   &  $3.4 (b^L) \times 10^{-3}$   &   $6.4(b^L)^2 \times  10^{-4}$ \\
\hline\hline
 $Y^D$   &    1   &   $3.4 (b^D) \times 10^{-3}$   &    $2.3(b^D)^2 \times 10^{-5}$ \\
\hline
\end{tabular}
\end{center}
 \caption{\small Hierarchies of fermion masses from the flux expansion.
Second order in derivatives.}
\label{jerarsm2}
\end{table}
Note that here the $Y_{22}$ entries have the same structure because 
in the three cases $Y_R+Y_L=\pm3$ (we are ignoring the overall sign of
the contribution which is not relevant for this estimate). As expected, the corrections 
to the Yukawa couplings are always higher than those coming from only first derivatives.
This is true even for the leptons, which have the highest maximal hypercharge and hence get
the largest contribution to first order in derivatives.
In fact, to avoid too large $Y_{11}^{U,L}$ values one rather needs $b^U,b^L <  1$.
On the other hand, the contribution to  $Y_{22}^L$ is still too small. The same happens with
the $D$-quarks, one would need  $b^D\simeq 5.6$ to reproduce the observed $D$-quark 
mass hierarchies, so that strong variation is again required for $D$-quarks.
Terms of order 3 and 4 in flux derivatives could also add to the relatively
large values of the $D$-quarks. Equation (\ref{yukder411}) shows that the expected contribution is
of order $Y_{11}\simeq \e\eta^4(d\hat q+d'\hat q') \simeq 2.6 d \times 10^{-4}$, which reproduces 
the $D$-quark mass result for a flux parameter $d\simeq 2.4$. Hence, if we do not want to
rely on relatively large flux parameters the case of $D$-quarks requires substantial
input from higher orders in the derivative expansion, up to order four.

In \cite{hv3} it was pointed out that the flux
expansion to first order in derivatives gives a good explanation of the
hierarchies observed for leptons and $U$-quarks but  terms coming from the higher
derivative flux expansion were  needed in order to describe the hierarchies 
for $D$-quarks. It was also suggested that a possible reason for this different 
behavior could arise from the fact that leptons and $U$-quarks have higher 
maximal hypercharge than the $D$-quarks. We indeed find that the hierarchies for 
$U$-quarks may be quite well described by  first order flux variations 
of order one.  The resulting electron mass is also of the correct order.
However, the dependence on hypercharge does not seem to explain
the different behavior of $L$ and $U$ compared to $D$ fermions. In particular, higher derivative terms always 
generically dominate over the first order terms, even taking into account the hypercharge dependence. 
The milder behavior  of the $D$-quark hierarchies can be  understood either by assuming a 
relatively strong first/second order flux variation (i.e. $a^D\simeq 5.8$ or  $b^D\simeq 5.6$)
or larger 4th order contributions with $d\simeq 2.4$. The muon has the tendency to come 
out too light which may indicate that neglecting flux variation in the Higgs matter curve
could perhaps be inappropriate for the leptons and possibly for the $D$-quark matter curves.

If we assume that $U$-quarks get their Yukawas already at first 
order in derivatives (eq.(\ref{flx2})) and on the contrary the
$D$-quarks need a dominant  contribution at order two or higher 
(eq.(\ref{der22}) or eq.(\ref{DER}), it does not matter for this approximation), we can
also give an estimate of the CKM mixing matrix \cite{hv3}. Indeed in this case 
the respective mass squared matrices will be proportional to
\beq
Y^{U}(Y^U)^{\dagger} \ \sim \
\left(
\begin{array}{ccc}
 \e^4\eta^4  &  \e^3\eta^3    &  \e^2 \eta^2 \\
 \e^3 \eta^3 & \e^2 \eta^2   &  \e \eta \\
\e^2 \eta^2  &  \e \eta  &  1
\end{array}
\right)\  ;
\ 
Y^{D}(Y^D)^{\dagger} \ \sim \
\left(
\begin{array}{ccc}
 \e^2\eta^4  &  \e^2\eta^3    &  \e \eta^2 \\
 \e^2 \eta^3 & \e^2 \eta^2   &  \e \eta \\
\e  \eta^2  &  \e \eta  &  1
\end{array}
\right)
\label{upydown}
\eeq
Then, as in \cite{fn}, one can estimate the matrices $V^{U,D}$ which diagonalize each 
of them 
\beq
V^U \ \sim \
\left(
\begin{array}{ccc}
  1  &     \e \eta    &  \e^2 \eta^2 \\
 \e \eta  &  1    &  \e \eta \\
\e^2 \eta^2  &  \e \eta  &  1
\end{array}
\right)\  ;
\
V^D \ \sim \
\left(
\begin{array}{ccc}
  1  &  \eta    &  \e \eta^2 \\
  \eta   &   1     &  \e \eta \\
\e  \eta^2  &  \e \eta  &  1
\end{array}
\right)
\eeq
The CKM matrix, $V^{CKM} \simeq  V^U(V^D)^{\dagger}$, then turns out to be
\beqa 
V^{CKM} 
& \simeq & 
\left(
\begin{array}{ccc}
  1  &     \eta    &  \e \eta^2 \\
  \eta  &  1    &  \e \eta \\
\e  \eta^2  &  \e \eta  &  1
\end{array}
\right)
\ \simeq \ 
\left(
\begin{array}{ccc}
  1  &     \alpha_G^{1/2}    &  2\pi \alpha_Y^{1/2}\alpha_G \\
  \alpha_G^{1/2}   &  1    &  2\pi \alpha_Y^{1/2} \alpha_G^{1/2} \\
2\pi \alpha_Y^{1/2}\alpha_G  &  2\pi \alpha_Y^{1/2}\alpha_G^{1/2}  &  1
\end{array}
\right)
\nonumber \\[4mm]
\hspace*{2cm} & \simeq &
\left(
\begin{array}{ccc}
  1  &     0.20    &   0.006 \\
  0.20   &  1    &    0.03 \\
  0.006   &   0.03   &  1
\end{array}
\right)
\label{CKM}
\eeqa
which is in reasonable  agreement with experiment.
This structure is similar to that found in \cite{hv3}, although in comparison,
in the above formula the separate dependence on the hypercharge
flux is explicit and the 3rd generation mixing is slightly smaller.

As a general conclusion,  in this simplified scheme in which we have set the
flux variation in the third curve to zero, one can  reproduce the 
general pattern of quark and lepton hierarchies as well as quark mixing,
for reasonable choices of flux variation parameters. This is particularly
the case for the $U$-quarks and the electron. Nevertheless, a more complete
numerical study, not neglecting flux parameters of the Higgs matter curve,
may be required to get full agreement. The order of magnitude estimates
for the CKM matrix are on the other hand quite promising.
We leave a  more detailed  phenomenological analysis of this framework for future work.

\section{Final comments}

In this paper we have studied the local structure of zero mode wave functions of chiral matter fields in F-theory compactifications. 
We have solved the relevant differential equations for the zero modes which were derived from local Higgssing in the world-volume 
effective action of the F-theory 7-branes \cite{bhv1}. These wave functions have a Gaussian  profile centered on the matter 
curves and become distorted in the presence  of $U(1)$ fluxes both on the bulk and on the matter curves themselves. In our approach 
we first write the fluxes in a power series of the local coordinates and then make a perturbative expansion of the wave functions in 
powers of the flux coefficients. In this way we obtain expressions which may then be applied to compute physical quantities of interest. 
In this paper we have concentrated on the calculation of Yukawa couplings but the wave functions could also help to examine other
problems. For instance, they could be used to explore the effects of closed string fluxes and warping on the effective action, 
which could prove important in relation to compactifications with broken supersymmetry.

With the wave functions at our disposal we have computed Yukawa couplings by
performing explicitly the overlap integrals of the three wave functions 
linked to fermions and the Higgs field. By choosing an appropriate  gauge, 
the wave functions of quark or lepton generations are shown to
depend only on the bulk fluxes but not on the extra $U(1)$'s associated to 
the unfolding of the singularities. For example, in the case of a 
$SU(5)$ F-theory GUT  broken to the SM by hypercharge flux, the
effective distortion of the wave function depends on the hypercharge 
of the specific particle considered. The Yukawa integrals 
can be done analytically and in appendix B we provide the leading terms 
in the flux expansion. One interesting fact we find is that for a
constant non-localized Higgs wave function, presumably corresponding to a
Higgs field living on the bulk of the base $S$, the  flux distortion  
cancels in such a way that the possible Yukawa matrices remain of rank 
one. On the other hand, when the three wave functions are 
localized, corresponding  to three intersecting matter curves,
a non-constant $U(1)$ flux gives rise naturally to a hierarchy 
of Yukawa couplings as first pointed out in \cite{hv3}.

We have applied our findings to the understanding of the observed hierarchies of quark and lepton masses and mixings. In a 
simplified situation in which the flux variation in the Higgs matter curve is
negligible we obtain explicit compact formulas for  Yukawa couplings 
as a function of flux parameters and the charges of the bulk $U(1)$. 
In a $SU(5)$ setting broken to the SM by hypercharge flux, the resulting Yukawa couplings depend on different powers of the 
hypercharge of each quark and lepton. It turns out that reasonable values of flux parameters, involving only a first derivative
expansion of the fluxes, can account for the hierarchical structure of the masses of $U$-quarks and 
the electron.
The explanation of $D$-quark hierarchies seems to require larger contributions  from the higher 
order terms in the flux derivative expansion.
A reasonable  semiquantitative understanding of the CKM matrix is then obtained somewhat analogous to the results in
\cite{hv3}. 

The natural appearance of  hierarchies for masses and mixings looks
quite promising. However, a full explanation of the data 
would require a more detailed phenomenological analysis. 
In particular in the numerical estimations we assumed weakly varying fluxes
in the Higgs matter curve, which needs not necessarily be the case.
Furthermore, we also took flux variations of the same order for the matter 
curves corresponding to left- and right-handed fermions, which again
is suggestive but not generally true.
We think that our explicit formulas are a good starting point for a 
more thorough investigation which we plan to carry out elsewhere
\cite{afi}. 

Another interesting topic to address is the 
origin and structure of neutrino masses, which seem to follow a pattern
quite distinct from that of quarks. Here the crucial point is the 
nature and origin of the mass of right-handed neutrinos.
We think that our results will also be useful in this case.
More generally,  $U(1)$ fluxes may have meaningful implications
for other physical issues such as supersymmetry breaking. As an example,
in \cite{aci} it was proposed that in F-theory or type IIB orientifolds, 
local volume modulus dominance of supersymmetry breaking gives rise to
a very predictive  pattern of soft terms
consistent with radiative electroweak symmetry breaking. It was 
also pointed out that the presence of $U(1)$ fluxes affects in a
small but significant way the values of the soft terms and that 
these flux contributions could be needed in fact in order to obtain the
proper amount of neutralino dark matter. 
Corrections coming from hypercharge fluxes could also play an
important role in the detailed understanding of gauge coupling unification
\cite{dw1,blumen}. It thus appears that 
the distortion caused by fluxes could be indeed important in several physical issues 
in F-theory unification.

\vspace*{2cm}

\noindent
{\bf Acknowledgments}\\
We thank  F. Marchesano, A. Uranga and S. Theisen for useful advice.
A.F. acknowledges a research grant No. PI-03-007127-2008 from CDCH-UCV, as well
as hospitality and support from the Instituto de F\'{\i}sica Te\'orica UAM/CSIC, and the Max-Planck-Institut f\"ur Gravitationsphysik,
during completion of this paper. L.E.I. thanks the PH-TH Division at CERN for hospitality
while writing up this paper.
This work has been supported  by the CICYT (Spain) under project
FPA2006-01105, the Comunidad de Madrid under project HEPHACOS
P-ESP-00346 and the Ingenio 2010 CONSOLIDER program CPAN.

\clearpage

\noindent
{\bf \large Note added}

The Yukawa couplings among fields on curves $\Sigma_a$, $\Sigma_b$ and $\Sigma_c$ arise from the
superpotential term
\beq
W_Y   =   M_*^4\int_{ S} \Tr( {\pmb A_a} \wedge {\pmb A_b}\wedge {\pmb \Phi_c}) \ + \ {\rm cyclic \ permutations}
\label{superpot2}
\eeq
where $\pmb A$ and $\pmb \Phi$ are chiral superfields given in (\ref{superfi}).
It is enough to focus on $\theta\theta$ terms involving two fermions and one scalar. 
The three families of quark and leptons are taken to reside in curves $\Sigma_a$ and $\Sigma_b$ 
while the Higgs lives on $\Sigma_c$. 
Then, neglecting an overall constant, the coupling is given by
\beqa
\tilde Y_{ij} = \int_S \big[\psi^i_{a\bar 1} \psi^j_{b\bar 2} \varphi_c & - & \psi^i_{a\bar 2} \psi^j_{b\bar 1} \varphi_c 
\ + \ \psi_{c\bar 1} \psi^i_{a\bar 2} \varphi_b^j 
\label{fullyuk}\\[2mm]
& - & \psi_{c\bar 2} \psi^i_{a\bar 1} \varphi_b^j  \ + \
\psi^j_{b\bar 1} \psi_{c\bar 2} \varphi^i_a \ - \  \psi^j_{b\bar 2} \psi_{c\bar 1} \varphi^i_a \big]
\mathsm{dz_1 \wedge d\bar z_1 \wedge dz_2 \wedge d\bar z_2}
\nonumber
\eeqa
Note that the Yukawa computations in the main text of the paper involve only the contribution
from the first two terms. On the other hand, in a fully symmetric
local interaction the additional four terms from cyclic permutations should also be included.
This has been recently addressed in refs.\cite{cchv} and \cite{cp}.
In \cite{cchv} it has been shown that $\tilde Y_{ij}$ does not receive corrections when fluxes are turned on.
We wish to stress that there is a delicate  cancellation 
among the six contributions in eq.(\ref{fullyuk}), each term being in general 
flux dependent. This happens independently
of whether or not the field strengths satisfy the BPS condition $\omega \wedge F =0$.

It is instructive to consider the example of constant fluxes. In this case it can be exactly shown 
that each non-trivial term in  
(\ref{fullyuk}) separately gives a flux dependent contribution to the third generation coupling $\tilde Y_{33}$, but the 
full coupling is flux independent. As in \cite{cchv}, using our notation, we turn a gauge field along the $Q_1$ and $Q_2$
directions given by $A=A_a Q_1 + A_b Q_2$. Furthermore, we choose
\beq
A_a=A_b=-iM \bar z_1 dz_1 - iN \bar z_2 dz_2 + {\rm c.c.}
\label{cfluxab}
\eeq
Notice that the gauge field acting on $\Sigma_c$ is $A_c=-(A_a+A_b)$. The resulting field strength satisfies the BPS
condition provided $M+N=0$.

The zero modes on each curve follow from the results in appendix A. We find
\beqa
\Sigma_a & : & \varphi_a = f_a(z_2) e^{-\lambda_a |z_1|^2} \quad ; \quad
\lambda_a=-M + \frac1{v}\sqrt{1+ M^2 v^2} 
\nonumber \\[2mm]
\Sigma_b & : & \varphi_b=  f_b(z_1) e^{-\lambda_b |z_2|^2} \quad ; \quad
\lambda_b=-N + \frac1{v}\sqrt{1+ N^2 v^2} 
\label{zmcf} \\[2mm]
\Sigma_c & : & \varphi_c = e^{-\lambda_c |w|^2} e^{\xi w\bar u} 
\quad ; \quad \textstyle{2v^2\lambda_c(\lambda_c-2M)(\lambda_c-2N) - (\lambda_c-M-N)=0}
\nonumber 
\eeqa
where $\xi= \frac{\lambda_c(N-M)}{(\lambda_c-M-N)}$. Here we have already set $f_c=1$ in the curve $\Sigma_c$ that
is taken to host the Higgs. Also, we will take $f_a(z_2)=z_2^{3-i}$ and $f_b(z_1)=z_1^{3-j}$. Since we are
working in the holomorphic gauge, from the zero mode equations (\ref{hatflux}), we further have
$\psi_{a\bar\jmath}=v\bar\partial_{\bar\jmath}\varphi_a/z_1$, 
$\psi_{b\bar\jmath}=v\bar\partial_{\bar\jmath}\varphi_b/z_2$ and 
$\psi_{c\bar\jmath}=-v\bar\partial_{\bar\jmath}\varphi_c/w$, 
where we have dropped the family index to ease notation. Observe that
in the example of constant fluxes these expressions
lead to simple results such as $\psi_{a\bar 1}=-v\lambda_a \varphi_a$, $\psi_{a\bar 2}=0$, and so on, so that 
only three of the terms in (\ref{fullyuk}) are not zero. It is straightforward to show that the coupling
vanishes except when $i=j=3$, and that
\beq
\tilde Y_{33}= v^2\big[\lambda_a\lambda_b + \lambda_a(\lambda_c + \xi) + \lambda_b(\lambda_c - \xi)\big]
\int_S \! d^2z_1 d^2 z_2 \, \varphi_a \varphi_b \varphi_c
\label{ty33}
\eeq
where the $\varphi_I$ are given in (\ref{zmcf}). Evaluation of the Gaussian integral yields
\beq
\int_S \!\! d^2z_1 d^2 z_2 \, \varphi_a \varphi_b \varphi_c = 
\frac{\pi^2}{\lambda_a\lambda_b + \lambda_c(\lambda_a + \lambda_b) + \xi(\lambda_a-\lambda_b)}
\label{iy33}
\eeq
Therefore, $\tilde Y_{33}=\pi^2v^2$, independent of fluxes. However, notice that each separate term
in (\ref{ty33}) depends on fluxes even if the BPS condition $M+N=0$ is satisfied.

In the example of \cite{cchv}, in which the BPS condition $\omega \wedge F=0$ is satisfied,
it also happens that the flux effects on the couplings only cancel when all terms in (\ref{ty33}) are included.
On the other hand, in the setup of this article, in which $\omega \wedge F=0$ is not enforced, nonetheless
it can be checked that when all terms in  (\ref{ty33}) are added only the coupling $\tilde Y_{33}$ survives
and is flux independent. In \cite{cp} the sum of all contributions to the couplings has also been taken
into account.

In \cite{cchv} it was proved that the cancellation of flux effects in the full coupling $\tilde Y_{ij}$
follows from an exact residue formula. For a pedestrian derivation of this formula we start from 
(\ref{fullyuk}) and manipulate the integrand to write it as a sum of total derivatives. To this purpose,
following \cite{cchv}, we write the zero modes $\psi_{I\bar \jmath}$ and $\varphi_I$, which satisfy
the last two equations in (\ref{hatflux}), as $\psi_{I\bar \jmath}=\bar\partial_{\bar \jmath}\, \xi_I$, together with
\beq
\varphi_a = \frac{z_1}{v} \xi_a + h_a \quad ; \quad
\varphi_b = \frac{z_2}{v} \xi_b + h_b \quad ; \quad 
\varphi_c = -\frac{w}{v} \xi_c + h_c 
\label{phixi}
\eeq
The functions $h_I$ are holomorphic and correspond to $\varphi_I\big|_{\Sigma_I}$.
An elementary calculation, dropping family indices to simplify, then shows that  
\beqa
\tilde Y \!\! & = & \!\!\int_S\!\!  d^2z_1 d^2 z_2 \, \left\{
\bar\partial_{\bar 1}\left[(\psi_{a\bar 2} \varphi_b -  \psi_{b\bar 2} \varphi_a) \xi_c\right]
- \bar\partial_{\bar 2}\left[(\psi_{a\bar 1} \varphi_b -  \psi_{b\bar 1} \varphi_a) \xi_c\right] \right\}
\nonumber\\[2mm]
& + & 
\int_S\!\!  d^2z_1 d^2 z_2 \, \left\{
\bar\partial_{\bar 1}\left( h_c\, \xi_a \bar\partial_{\bar 2} \xi_b \right)
- \bar\partial_{\bar 2}\left(h_c\, \xi_a \bar\partial_{\bar 1} \xi_b \right) \right\}
\label{presi}
\eeqa
The integrals in the first line can be evaluated by parts, and then the boundary terms are seen
to vanish because the zero modes $\varphi_a$ and $\varphi_b$ are localized. In the second line,
integrating by parts twice, using $\xi_a=v(\varphi_a - h_a)/z_1$, similarly for $\xi_b$, and 
invoking localization, gives the final residue formula 
$\tilde Y \sim {\rm Res}\left(\frac{h_a h_b h_c}{z_1z_2}\right)$
\cite{cchv}.

The computation of Yukawa couplings just described is purely local. 
If the symmetry among the cyclic permutations in eq.(\ref{fullyuk})
still remains after a global completion of the theory, only one generation
acquires a Yukawa coupling. In this case the observed hierarchy of fermion masses
cannot be generated just by turning on magnetic fluxes, some additional ingredient,
e.g. non-perturbative effects, should also be at work to produce these mass hierarchies.

\clearpage

\appendix
\section{Fluxed zero modes and wave functions}
\label{a:gensols}

In this appendix we study the solutions of the zero mode equations (\ref{withflux}) both for constant and variable 
field strengths. We will explicitly consider the curves $\Sigma_a$ and $\Sigma_c$. As in the fluxless case,
the results for $\Sigma_a$ and $\Sigma_b$ are completely similar, but the curve $\Sigma_c$ must be treated separately.

We find it convenient to rewrite the total gauge potential as
\beq
\ca = \hat \ca + d\Omega
\label{gtrans1}
\eeq
in such a way that $\hat \ca_{\bar 1}=\hat \ca_{\bar 2}=0$.
We can then work in this
`holomorphic'  gauge where the potential is just $\hat \ca$ and the corresponding fermions are denoted
$\hat \chi$ and $\hat \psi$. The advantage is that the equations reduce to
\beqa
(\partial_2 - i\hat \ca_2) \hat\psi_{\bar 2} + (\partial_1 - i\hat \ca_1)\hat\psi_{\bar 1} - 
m^2(\bz_1 q_1 + \bz_2 q_2) \, \hat\chi & = & 0 \nonumber \\[1mm]
\bar \partial_{\bar 1} \hat \chi - m^2(z_1 q_1 + z_2 q_2) \, \hat\psi_{\bar 1} & = & 0 \label{hatflux} \\[1mm]
\bar \partial_{\bar 2} \hat \chi - m^2(z_1 q_1 + z_2 q_2) \, \hat\psi_{\bar 2} & = & 0 \nonumber
\eeqa    
and the gauge fields do not appear in the last two equations.
The further constraint $\bar\partial_{\ca} \psi=0$ becomes $\bar\partial \hat\psi=0$ and is automatically
verified on account of the last two equations above.

The solutions for the original flux are recovered by performing a gauge transformation, namely
\beq
\chi = e^{i\Omega}\hat \chi \qquad ; \qquad \psi_{\bar 1} = e^{i\Omega}\hat \psi_{\bar 1} \qquad ; \qquad 
\psi_{\bar 2} = e^{i\Omega}\hat \psi_{\bar 2}
\label{osols}
\eeq
To compute Yukawa couplings it suffices to work with the hatted fields because the couplings are gauge invariant.

\subsection{Constant flux}
\label{a:cflux}

{}From the total gauge potential given in (\ref{tpotc}) it follows that
the transformed potential and gauge function are 
\beq
\hat \ca  = -2i\cam \bz_1 dz_1 \, - \, 2i\cn \bz_2 dz_2 \qquad ; \qquad \Omega=i(\cam |z_1|^2 + \cn |z_2|^2) 
\label{tpotcnew}
\eeq
We then need to find the solutions of (\ref{hatflux}) when $\hat \ca_1=-2i\cam \bz_1$ and $\hat \ca_2=-2i\cn \bz_2$.
The charges $(q_1, q_2)$ that must also be specified depend on the curve. 

\bigskip
\noindent
\underline{$\Sigma_a$, $(q_1,q_2)=(e_1,0)$}

\medskip
\noindent
Notice that in this case $\cam=qM$ and $\cn=qN+e_1N^{(1)}$, where $(M,N)$ come from the bulk flux and $N^{(1)}$ from
the flux along the curve. As in the fluxless case, we find that $\hat \psi_{\bar 2}=0$, which then implies
$\bar \partial_{\bar 2} \hat \chi=0$. We make the Ansatz
\beq
\hat \chi = f(z_2)\, e^{-\lambda_1 |z_1|^2}
\label{hatchi1}
\eeq 
The equation $\bar \partial_{\bar 1}\hat \chi=z_1\hat \psi_{\bar 1}$ then fixes
\beq
\hat \psi_{\bar 1} = - \frac{\lambda_1}{e_1m^2} \hat \chi
\label{hatpsi1}
\eeq
There is still an equation that requires $\lambda_1$ to satisfy
\beq
\lambda_1^2 + 2\cam \lambda_1 - e_1^2m^4 = 0
\label{eql1}
\eeq 
To have localized solutions we choose the root
\beq
\lambda_1 = -\cam + e_1m^2\, \sqrt{1 + \frac{\cam^2}{e_1^2m^4}} 
\label{l1exp}
\eeq 
which reduces to $\lambda_1=e_1 m^2$ when $\cam =0$.
Inserting in (\ref{hatchi1}) and (\ref{hatpsi1}) gives the solutions found in \cite{hv3} in a different gauge.

\bigskip
\noindent
\underline{$\Sigma_c$, $(q_1,q_2)=(-e_1,-e_2)$}

\medskip
\noindent
In the fluxless case we saw that to solve the equations it is convenient to set $e_1=e_2=e$, 
and to use the variables $w=(z_1+z_2)$ and $u=(z_1-z_2)$, together
with the redefined fermions $\psi_{\bar w}=(\psi_{\bar 1} + \psi_{\bar 2})/2$, and 
$\psi_{\bar u}=(\psi_{\bar 1} - \psi_{\bar 2})/2$.
 
The gauge potential is still formally given by (\ref{tpotcnew}) but now $\cam=(qM -e_2M^{(2)})$ and $\cn=(qN-e_1N^{(1)})$.
In the new variables the non-vanishing components of $\hat \ca$ are
\beqa
\hat \ca_w & = & - i \Delta \bar w \, - \, i(\cam-\Delta) \bar u \nonumber \\[1mm]
\hat \ca_u & = & - i  \Delta \bar u \, - \, i(\cam-\Delta) \bar w
\label{hpot3}
\eeqa
where $\Delta=(\cam+\cn)/2$. In the gauge $\hat \ca$, the zero mode equations imply that the
$\hat \psi$ fermions neatly depend on $\hat \chi$ as
\beq
\hat \psi_{\bar w} = - \frac1{em^2w} \, \bar \partial_{\bar w} \hat \chi \qquad ; \qquad
\hat \psi_{\bar u} = - \frac1{em^2w} \, \bar \partial_{\bar u} \hat \chi   
\label{psichi}
\eeq
In turn $\hat \chi$ can be determined from the remaining equation
\beq
(\partial_w - i\hat \ca_w) \hat\psi_{\bar w} \, + \, (\partial_u - i\hat \ca_u)\hat\psi_{\bar u} \, + \, 
\frac12 e m^2\, \bar w \hat\chi =  0
\label{hatchieq}
\eeq
To solve we make the Ansatz
\beq
\hat \chi = h(u+\gamma w) \, e^{-\lambda_3 |w|^2}\, e^{\xi w \bar u}
\label{chiansatz}
\eeq
where $h(u+\gamma w)$ is a holomorphic function of its argument. It then follows
\beq
\hat \psi_{\bar w} = \frac{\lambda_3}{em^2} \hat \chi \qquad ; \qquad
\hat \psi_{\bar u} = - \frac{\xi}{em^2} \hat \chi   
\label{psichi2}
\eeq
Substituting in (\ref{hatchieq}) determines the unknown constants. We find
\beq
\gamma=\frac{\xi}{\lambda_3} \qquad ; \qquad \xi=\frac{\lambda_3(\cam-\Delta)}{(\lambda_3 + \Delta)}
\label{xieps}
\eeq
Finally, $\lambda_3$ is a positive root of the cubic equation
\beq
\lambda_3(\lambda_3 + \cam)(\lambda_3 + 2\Delta-\cam) \, - \, 
\frac12 e^2 m^4 (\lambda_3 + \Delta)  = 0 
\label{eql3}
\eeq
When $\cam=\cn=0$ we recover the fluxless solution with $\xi=\gamma=0$, and $\lambda_3=em^2/\sqrt{2}$.
In the special cases $\Delta=0$ ($\cn=-\cam$) and  $\Delta=\cam$ ($\cn=\cam$) the cubic becomes quadratic
and the positive root is easily identified. 

\subsection{Variable flux}
\label{a:vflux}

We consider the quadratic flux given in (\ref{vflux}). The corresponding transformed potential and the 
gauge function turn out to be
\beqa
\hat \ca_1 & = &  -2i\cam \bz_1 \, - \, 2i(\bar\a_1\bz_1^2 + 2\a_1z_1\bz_1) \, - \, 
2i(\bar\b_1 \bz_1^3 + 3\b_1\bz_1 z_1^2) \nonumber\\[1mm]
\hat \ca_2 & = &  -2i\cn \bz_2 \, - \, 2i(\bar\a_2\bz_2^2 + 2\a_2z_2\bz_2) \, - \, 
2i(\bar\b_2 \bz_2^3 + 3\b_2\bz_2 z_2^2)  
\label{vpothat} \\[1mm]
\Omega & = & i|z_1|^2\left[\cam + i(\a_1 z_1 + \bar\a_1 \bz_1) \, + \, 
(\b_1 z_1^2 + \bar\b_1 \bz_1^2) \right]  \nonumber \\[1mm] 
& + & i|z_2|^2\left[\cn \, + \, (\a_2 z_2 + \bar\a_2 \bz_2) \, + \, (\b_2 z_2^2 + \bar\b_2 \bz_2^2)\right] \nonumber  
\eeqa
As described below for particular curves, we have only been able to obtain zero mode solutions in a
perturbative expansion in the flux parameters.

\bigskip
\noindent
\underline{$\Sigma_a$, $(q_1,q_2)=(e_1,0)$}

\medskip
\noindent
As in the constant flux case we find $\hat \psi_{\bar 2}=0$ which implies $\bar \partial_{\bar 2} \hat \chi=0$. 
On the other hand, 
$\hat \psi_{\bar 1}=\frac{v}{z_1} \bar\partial_{\bar 1}\hat\chi$, where $v=1/e_1m^2$. There is still an equation  
\beq
(\partial_1 - i\hat \ca_1) \hat\psi_{\bar 1} \, - \, e_1 m^2 \bar z_1\, \hat\chi =  0
\label{hatchieq2}
\eeq
with $\hat \ca_1$ given in (\ref{vpothat}). We have found a solution $\hat\chi = \sum_{I=0} \hat \chi^{(I)}$,
where $\hat  \chi^{(I)}$ is of order $I$ in the flux coefficients. There is a corresponding expansion for
$\hat \psi_{\bar 1}$ with $\hat \psi_{\bar 1}^{(I)}=\frac{v}{z_1} \bar\partial_{\bar 1}\hat\chi^{(I)}$.

The zeroth order solutions are the fluxless ones presented in section \ref{ss:noflux}. They are
\beq
\hat \chi^{(0)} = f(z_2) \, e^{-e_1 m^2|z_1|^2} \qquad ; \qquad 
\hat \psi_{\bar 1}^{(0)} = - f(z_2) \, e^{-e_1 m^2|z_1|^2}  
\label{s1zeroth}
\eeq
The expansion of $\hat\chi$ to second order turns out to be
\beq
\hat \chi =  \hat \chi^{(0)} \bigg\{ 1  + v^3 H_{23} + v^2 H_{22} +  v (H_{21} + H_{11}) 
+ H_{10} + \frac12 H_{10}^2 + \cdots \bigg\}
\label{chi12} 
\eeq
where $v=1/e_1m^2$ is the volume defined before. The auxiliary functions are given by
\beqa
H_{10} & = &  z_1 \bz_1\left[\cam + \frac23(\bar\a_1 \bz_1 + 2\a_1 z_1) 
+ \frac12(\bar\b_1 \bz_1^2 + 3\b_1 z_1^2)\right] 
\label{h10def}  \\[2mm]
H_{11} & = &  \frac43 \a_1 z_1 + \frac32\b_1 z_1^2  
\label{h11def} \\[2mm]
H_{21} & = & -z_1\bz_1\bigg[\frac12\cam^2  + \frac49\cam(2\bar\a_1\bz_1 - \a_1z_1) - (\frac43\a_1^2+\frac34\cam\b_1)z_1^2
+(\frac34\cam\bar\b_1+\frac49\bar\a_1^2)\bz_1^2  \nonumber\\[1mm]
&& \hspace*{-5mm}
-\, \frac{16}5\a_1\b_1 z_1^3 + \frac45\bar\a_1\bar\b_1\bz_1^3 - \frac15 \bar\a_1\b_1 z_1^2\bz_1
+\frac2{15}\a_1\bar\b_1 z_1\bz_1^2  -\frac{15}8\b_1^2 z_1^4 + \frac38\bar\b_1^2\bz_1^4\bigg]
\label{h21def} \\[2mm]
H_{22} & = & z_1\left[\frac49\cam\a_1+ (\frac43\a_1^2+\frac34\cam\b_1)z_1 + \frac{16}5\a_1\b_1z_1^2+\frac{15}8\b_1^2z_1^3\right]
\nonumber \\[1mm]
& & \hspace*{10mm} 
+ \, \frac2{15}z_1\bz_1 \left(2\bar\a_1\b_1 z_1 - \a_1\bar\b_1 \bz_1 \right)
\label{h22def} \\[2mm]
H_{23} & = & \frac4{15}\bar\a_1\b_1 z_1 
\label{h23def} \
\eeqa
It can be checked that when $\a_1=\b_1=0$, the results match those of section \ref{a:cflux}
to second order in $\cam$.

The expansion of the wave function $\hat \psi_{\bar 1}$ needed to compute Yukawa couplings 
follows from  $\hat \psi_{\bar 1}=\frac{v}{z_1} \bar\partial_{\bar 1}\hat\chi$. We obtain
\beq
\hat \psi_{\bar 1} =  \hat  \psi_{\bar 1}^{(0)} 
\bigg\{ 1 + v^3 H_{23}^* + v^2 K_{22} -  v(\cam + H_{11}^* + K_{21}) 
+ H_{10} + \frac12 H_{10}^2 + \cdots \bigg\}
\label{psi12}
\eeq
with the additional definitions
\beqa
K_{22} & = & \frac12\cam^2  + \bz_1\bigg[\frac{16}9\cam \bar\a_1 
+(\frac43\bar\a_1^2 + \frac94\cam\bar\b_1)\bz_1 + \frac{16}5\bar\a_1\bar\b_1\bz_1^2  + \frac{15}8\bar\b_1^2 \bz_1^3\bigg] 
   \nonumber\\[1mm]
&& \hspace*{10mm}
+ \, \frac2{15}z_1\bz_1 \left(2\a_1\bar\b_1 \bz_1 - \bar\a_1 \b_1 z_1 \right)
\label{k22def}\\[2mm]
K_{21} & = & z_1\bz_1\bigg[
\frac32\cam^2+\frac{2}9\cam(10\a_1z_1+13\bar\a_1\bz_1) +
(\frac49\a_1^2+\frac94\cam\b_1)z_1^2 +(\frac43\bar\a_1^2+\frac{11}4\cam\bar\b_1)\bz_1^2 
\nonumber \\[1mm]
&&
 \hspace*{5mm}
+ \, \frac83 |\a_1|^2 z_1\bz_1 
+\frac{14}5 \bar\a_1\b_1 z_1^2\bz_1
+\frac{14}{5}\a_1\bar\b_1 z_1\bz_1^2 + 3|\b_1|^2 z_1^2 \bz_1^2
\label{kdef}\\[1mm]
& & \hspace*{5mm}
+ \, \frac45\a_1\b_1z_1^3
+\frac{37}{15}\bar\a_1\bar\b_1\bz_1^3
+\frac38\b_1^2z_1^4
+\frac98\bar\b_1^2\bz_1^4 \bigg]
\nonumber
\eeqa

\bigskip
\noindent
\underline{$\Sigma_c$, $(q_1,q_2)=(-e_1,-e_2)$}

\medskip
\noindent
We need to solve the zero mode equations (\ref{psichi}) and (\ref{hatchieq}).
The gauge potential components $\hat \ca_w$ and $\hat \ca_u$ can be easily found  
changing to coordinates $w=(z_1+z_2)$ and $u=(z_1-z_2)$ starting from (\ref{vpothat}).  
As before we define $\Delta=(\cam+\cn)/2$. In analogy we also introduce 
\beq
\delta=\frac12(\a_1+\a_2) \quad ; \quad \rho=\frac12(\b_1+\b_2) 
\label{delabdefs}
\eeq
and the corresponding $\bar\delta = \delta^*$ and $\bar\rho=\rho^*$.

To iterate we begin with the zeroth order solutions presented in section \ref{ss:noflux}, taking $h=1$. They are
\beq
\hat \chi^{(0)} = e^{-em^2|w|^2/\sqrt{2}} \qquad ; \qquad 
\hat \psi_{\bar w}^{(0)} = \frac{1}{\sqrt{2}} \, e^{-em^2|w|^2/\sqrt{2}} \qquad ; \qquad
\hat \psi_{\bar u}^{(0)} = 0  
\label{s3zeroth}
\eeq
To higher orders we will only report the wave function $\hat\chi$ that enters in Yukawa couplings.
To first order we find
\beq
\hat \chi^{(1)} =  \hat \chi^{(0)} \bigg\{ \frac{\sqrt2}{em^2} D_{11} + D_{10} \bigg\}
\label{chi13} 
\eeq
with functional coefficients given by
\beqa
D_{10} \!\! & = & \!\! w\bigg[(\cam -\Delta) \bu+\frac12\,\Delta \bw + 
\frac12\left(\bar\a_1-\bar\delta\right) \bw\bu + \frac12\left(\alpha_{{1}}-\delta \right)\left(w\bu + u\bw\right)
+ \frac16 \bar\delta\left({\bw}^{2} + 3\bu^2 \right) 
\nonumber \\[1mm]
& & \hspace*{-0.8cm} 
+\frac13\,\delta\left({w}\bw + u \bu \right) 
+ \frac14\left(\bar\b_{{1}}-\bar\rho \right)\left( {\bu}^{3} + \bu\bw^2 \right)
+ \frac14\left(\beta_{{1}}-\rho \right) \left({w}^2\bu + 2 u w \bw + 3 u^2\bu \right)
\nonumber \\[1mm]
& & \hspace*{-0.8cm}
+ \frac1{16}\bar\rho \left({\bw}^{3}+ 6\bw{\bu}^{2}\right) 
+ \frac3{16}\rho \left(w^2\bw+ 4w u{\bu} + 2 \bw u^2 \right)  \bigg] 
\label{d10def} \\[2mm]
D_{11} & = & w\bigg[\frac43\,\delta\, + {\frac {9}{16}}\,\rho\,{w} + 2(\beta_1 -\rho) u \bigg]
\label{d11def}
\eeqa
We have also computed the second order correction to $\hat \chi$. We refrain from presenting it because
it involves too many terms.

\section{Yukawa couplings}
\label{b:yukresu}

The purpose of this appendix is to provide the explicit expressions for the Yukawa couplings $Y_{ij}$  obtained
upon performing the overlap integral of the localized wave functions on the curves $\Sigma_I$,
$I=a,b,c$. The procedure is to determine the integrand $z_2^{3-i} z_1^{3-j} G_a G_b G_c$, where the
$G_I$ are the corrections of the wave functions due to fluxes that were derived in appendix A.
The integral with measure (\ref{measure2}) is evaluated assuming that
the size of the compact manifold $S$ is much larger than the width $v$ of the matter curves.
The piece of the integrand that can contribute is a sum of terms $|w|^{2m}|u|^{2n}$
and the integral is easily computed.

\subsection{Flux expansion, first order in derivatives}
\label{b:f1}

To proceed systematically, to begin we consider the field strengths expanded up to linear order.
In this way we obtain Yukawa couplings depending only on 
the parameters $\a_I$ and $\d$ that characterize the first derivative of the total flux acting on 
the matter curves $\Sigma_I$. We focus on  the leading terms for each entry.
There are corrections proportional to powers of the zeroth order coefficients, $\cam$ and $\cn$,
times powers of the $\a_I$, which are always subleading, i.e. higher order in $\e$.
We have normalized with respect to the zeroth order third generation Yukawa coupling 
$Y_{33}^{(0)}=\pi^2 s$, where $s=\sqrt2-1$. The results are as follows:
\beqa
Y_{23} &  = & \frac{v^2}{R}\big[\, 0.11\ \aa -0.27\ \ab-0.41\ \ac+0.52\ \db \, \big] 
\label{y23a} \\[2mm]
Y_{22} &  =  & \frac{v^4}{R^2} \big[\, 0.067(\aa^2+\ab^2) +0.11\ \aa\ab -0.02\ \aa \ac \nonumber \\
 &  +  &
0.17\ \ac^2 + 0.02\ \ab \ac+1.11\ \db^2 -\db(0.33\ac+0.45\ab+0.42\aa)\, \big]
\label{y22a} \\[2mm]
Y_{13}  &  = & \frac{v^4}{R^2} \big[ -0.03\aa^2
-0.16\ab^2-0.32\ac^2 -0.04\aa \ab + 0.02 \aa \ac   \nonumber \\
&  -  & 0.30\ab \ac + 0.15\aa\db -1.83\db^2+ 1.21\ac\db + 0.97\ab \db \, \big] 
\label{y13a} 
\eeqa
\beqa
Y_{12} &  = & \frac{v^6}{R^3} \big[\, 0.03\aa^3+\aa^2(0.02\ab+0.01\ac-0.12\db) \nonumber  \\
&  + &
\aa(0.03\ab^2+0.01\ab\ac-0.03\ac^2-0.21\ab\db-1.08\ac\db+1.59\db^2) \nonumber  \\
&  -   &
4.32\db^3 +4.5\ac\db^2+1.67\ab\db^2-0.91\ac^2\db-1.5\ab\ac\db  \nonumber \\
&  +  &
0.12\ab^2\db+0.20\ac^3+0.08\ab\ac^2-0.06\ab^2\ac-0.07\ab^3 \, \big]
\label{y12a} \\[2mm]
Y_{11}  & = &  \frac{v^8}{R^4} \big[\, 
0.1(\aa^4+\ab^4)+\aa^3(0.09\ab-0.15\ac-0.53\db) + \aa^2(3.10\db^2-0.15\ac\db \nonumber \\
& -  & 1.03\ab\db+0.15\ac^2-0.04\ab\ac+0.12\ab^2)+\aa(-1.0\db^3-18.96\ac\db^2+
4.78\db^2\ab)  \nonumber \\
&  + &
\db\aa(8.40\ac^2-0.43\ab\ac-1.11\ab^2) +0.15\aa\ac^3 +0.22\aa\ab\ac^2 \nonumber \\
& + &
0.04\aa\ab^2\ac+0.09\ab^3\aa + 2.85\db^4+39.31\ac\db^3-4.07\ab\db^3 -
21.11\ac^2\db^2  \nonumber\\
&  -  & 
16.50\ab\ac\db^2 +3.4\ab^2\db^2 + 1.46\ac^3\db +9.33\ab\ac^2\db -0.45\ab^2\ac\db \nonumber \\
&  -  &
0.83\ab^3\db-0.36\ac^4-0.15\ab\ac^3+0.15\ab^2\ac^2+0.15\ab^3\ac  \, \big]
\label{y11a}
\eeqa
The couplings $Y_{ij}$ satisfy the property
\beq
Y_{ij}(\bar\a_a, \bar\a_b, \bar\a_c, \bar\d) = Y_{ji} (\bar\a_b, \bar\a_a, 2\bar\d - \bar\a_c, \bar\d)
\label{dualp}
\eeq
Then, the $Y_{ij}$ for $i > j$ can be easily found from the above results.

We want to stress that just as $Y_{23}$ given in (\ref{y23exact}), all couplings can be computed exactly. The results are given numerically only for
ease of presentation. For example, $Y_{22}$ is found to be
\beqa
Y_{22} & \! = \!& \frac{v^4}{3R^2}\big[\, s(s-1)(s^2-1)^2(\aa^2 + \ab^2) + \frac23 (s-1)^2(3s^2+1)\aa\ab 
\nonumber \\
& + & s(s-1)(3s-1)(\aa-\ab)\ac 
-  \frac34 \sqrt2 s(3s^2+4\sqrt2 s-4)\ac^2 
\label{y22salpha} \\
& + &  \frac14 s(16s^3-13\sqrt2 s^2 - 40s + 36\sqrt2 )\db^2 +\frac32 s(3\sqrt2s^2+ 8s-4\sqrt2) \db\ac
\nonumber \\
& +& \frac13 s (12s^3-15s^2+4s-9)\db\aa + \frac13 s (12s^3+3s^2-20s-3)\db\ab   
\big]
\nonumber
\eeqa 
where $s=\sqrt2-1$ is the parameter in the measure (\ref{measure2}).

For completeness we also provide the expansion of $Y_{33}$ to first order order in fluxes, namely
\beq
Y_{33} = 1 + v\big[\, \frac12(s-1)(\cam_a + \cn_b) + s \Delta \,\big] 
\label{y33exact}
\eeq
with $s=\sqrt2-1$. This is the simplest example showing that the corrections vanish when $\chi_c=1$
which implies $s=1$ and $\Delta=0$.

\subsection{Flux expansion, second order in derivatives}
\label{b:f2}

To second oder in derivatives of the fluxes there are further contributions
to the Yukawas with leading terms  as follows:
\beqa
\!\! Y_{13} & \!\!\! = \!\!\! & \frac{v^3}{R^2}\big[ -0.07\ba -0.44\bb-0.34\bc +1.01 \rb \, \big] 
\label{y13b}\\[2mm]
\!\! Y_{22} & \!\!\! = \!\!\! & \frac{v^3}{R^2} \big[ 0.18 (\ba + \bb) -0.57\rb \, \big]
\label{y22b}\\[2mm]
\!\! Y_{12} & \!\!\! = \!\!\!& \frac{v^5}{R^3} \big[ \ba (-0.16\aa-0.04\ab+0.04\ac+0.13\db)
+\bb(0.16\aa+0.39\ab+0.02\ac-0.61\db) \nonumber \\
& \!\!\! + \!\!\! &
\bc (0.91\aa+1.3\ab+0.36 \ac -4.13 \db) + \db(-0.81\aa -2.00\ab -1.7 \ac +6.29 \db \, \big] 
\label{y12b}\\[2mm]
\!\! Y_{11} &  \!\!\! =  \!\!\! & \frac{v^6}{R^4} \big[ 
-0.23(\ba^2+\bb^2)+ \ba (0.29\bb + 1.74 \bc -1.45\rb) \nonumber \\
& \!\!\! + \!\!\! &
\bb(2.03\rb - 1.74 \bc) -5.49\bc^2 +10.99 \bc \rb - 7.20 \rb^2 \, \big]
\label{y11b}
\eeqa
The coupling $Y_{31}$ follows from $Y_{13}$ by exchanging $\ba \leftrightarrow \bb$ and  $\bc \leftrightarrow (2\rb - \bc)$.
A similar remark applies to $Y_{21}$.

\subsection{Flux expansion, third and fourth orders in derivatives}
\label{b:f34}

In section \ref{ss:yukmat} we also discuss the effects of third and fourth order derivatives
in the fluxes. The modified wave functions needed to calculate the couplings are obtained as explained in \ref{a:vflux} 
but with new terms in the gauge potential because now the components of the total field strength have the additional pieces 
\beq
\cf_{i\bar \imath}^{extra}  =  8i(C_i z_i^3 + \bar C_i \bz_i^3) \, + \, 10i(D_i z_i^4 + \bar D_i \bz_i^4)
\label{vfluxcd}
\eeq
To third order the effective flux acting on the $\Sigma_I$ is characterized by parameters $C_I$ and
$\Delta_c$ as discussed in section \ref{ss:eflux}. The notation at fourth order is analogous. The
couplings that receive new corrections are  
\beqa
Y_{11} & \! = \!& \frac{5v^3}{4R^4}\big[\, (s-1)(s^2-1)^2(\bar D_a + \bar D_b) 
+ 2s(s^4-3s^2+3)\bar\Delta_d\, \big]
\label{y11exact} \\[2mm]
Y_{12} & \! = \! & \frac{v^4}{10R^3}\big[\, 6(s-1)^2(s^2-1) \bar C_a + 
6(s^2-1)^2 \bar C_b  -  15s(s^2-2) \bar C_d \nonumber \\ 
& + & s(12s^3+15s^2-20s-30)\bar\Delta_c \, \big]
\label{y12exact}\\[2mm]
\hspace*{-5mm}Y_{21} & \! = \! & \frac{v^4}{10R^3}\big[\, 6(s^2-1)^2 \bar C_a + 
6(s-1)^2(s^2-1) \bar C_b + 15s(s^2-2) \bar C_d \nonumber \\
& + & s(12s^3-15s^2-20s+30)\bar\Delta_c \, \big]
\label{y21exact}
\eeqa
where $s=\sqrt2-1$. Observe again that these couplings vanish when $\chi_c=1$.
Numeric evaluation gives 
\beqa
Y_{11} & = & \frac {v^5}{R^4} \big[-0.50(\Da + \Db) + 2.60 \bar\Delta_d) \, \big]
\label{y11c}\\[2mm]
Y_{12} & = & \frac {v^4}{R^3} \big[-0.17\Ca+0.41\Cb + 1.13\Cc-1.44\bar\Delta_c \, \big]
\label{y12c}\\[2mm]
Y_{21} & = & \frac {v^4}{R^3} \big[-0.17\Cb+0.41\Ca -  1.13\Cc+0.83\bar\Delta_c \, \big]
\label{y21c}
\eeqa

\end{document}